\documentclass[10pt, pre, amsmath, onecolumn, showpacs, superscriptaddress]{revtex4-1}
\usepackage{graphicx,color}
 \graphicspath{{./}{./figures/}}
\usepackage{amsmath}
\usepackage{enumerate}
\usepackage{mathbbol}
\usepackage{amsfonts}
\usepackage{natbib}

\usepackage{bm}

\usepackage{color}
\usepackage[dvipsnames]{xcolor}

\newcommand{\nn}{\nonumber}

\newcommand{\be}{\begin{equation}}
\newcommand{\ee}{\end{equation}}
\newcommand{\bea}{\begin{eqnarray}}
\newcommand{\eea}{\end{eqnarray}}
\newcommand{\beq}{\begin{eqnarray}}
\newcommand{\eeq}{\end{eqnarray}}

\newcommand{\bx}{{x}}

\newlength{\bilderlength}

\begin{document}

\title{Cumulants and large deviations for the linear statistics of the one-dimensional trapped Riesz gas}

\author{Pierre Le Doussal}
\affiliation{Laboratoire de Physique de l'Ecole Normale Sup\'erieure, PSL University, CNRS, Sorbonne Universit\'es, 24 rue Lhomond, 75231 Paris, France}
\author{Gr\'egory \surname{Schehr}}
\affiliation{Sorbonne Universit\'e, Laboratoire de Physique Th\'eorique et Hautes Energies, CNRS UMR 7589, 4 Place Jussieu, 75252 Paris Cedex 05, France}

\date{\today}

\date{\today}

\begin{abstract} 
We consider the classical trapped Riesz gas, i.e., $N$ particles at positions $x_i$ in one dimension with a repulsive power law interacting potential $\propto 1/|x_i-x_j|^{k}$, with $k>-2$, in an external confining potential of the form $V(x) \sim |x|^n$. We focus on the equilibrium Gibbs state of the gas, for which the density has a finite support $[-\ell_0/2,\ell_0/2]$.
We study the fluctuations of the linear statistics 
${\cal L}_N = \sum_{i=1}^N f(x_i)$ in the large $N$ limit for smooth functions $f(x)$. 
We obtain analytic formulae for the cumulants of ${\cal L}_N$ for general $k>-2$. 
For long range interactions, i.e. $k<1$, which include the log-gas ($k \to 0$) and the Coulomb gas ($k =-1$) 
these are obtained for monomials $f(x)= |x|^m$. 
For short range interactions, i.e. $k>1$, which include the Calogero-Moser model, i.e. $k=2$,
we compute the third cumulant of ${\cal L}_N$ for general $f(x)$ and arbitrary cumulants for monomials $f(x)= |x|^m$.
We also obtain the large deviation form of the probability distribution of ${\cal L}_N$, which
exhibits an ``evaporation transition'' where the fluctuation of ${\cal L}_N$ is dominated by the one of the largest $x_i$. 
In addition, in the short range case, we extend our results to a (non-smooth) indicator function $f(x)$,
obtaining thereby the higher order cumulants for the full counting statistics
of the number of particles in an interval $[-L/2,L/2]$. We show in particular that they exhibit an interesting scaling form as $L/2$ approaches
the edge of the gas $L/\ell_0 \to 1$,
which we relate to the large deviations of the emptiness probability of the complementary interval on the real line.
\end{abstract}

\maketitle

\section{Introduction}

\subsection{Overview} 

Linear statistics are a useful tool to characterize the fluctuations of systems of interacting particles. For a system of $N$ particles, 
they are defined as ${\cal L}_N = \sum_{i=1}^N f({x_i})$ where the $x_i$'s denote the positions of the particles and where $f(x)$ is an arbitrary function. For instance, for $f(x) = x$, the linear statistics ${\cal L}_N$ simply denotes the position of the center of mass of the gas of particles, but more general functions can be considered. The study of linear statistics amounts to determine the probability distribution of 
${\cal L}_N$ in particular in the limit of a large system $N \gg 1$. For independent and identically distributed (IID) random variables $x_i$,
assuming that $f(x_i)$ has first and second moments, the mean of ${\cal L}_N$ is of order $O(N)$ while the typical fluctuations are Gaussian and of order $O(\sqrt{N})$, thanks to the Central Limit Theorem (CLT). In the presence of interactions, the positions $x_i$'s become correlated, however 
in many cases the limiting
distribution of ${\cal L}_N$, upon proper rescaling, still converges to a Gaussian distributions. This leads to CLT's 
with a variance depending non-trivially on $f(x)$ and on the shape of the interaction \cite{JohanssonLS,Beenaker1963,RiderVirag2007,Serfaty2023}.
This variance, as well as higher order cumulants, contains coarse grained information about many-body correlations in the gas. 

In one spatial dimension $d=1$, linear statistics have been much studied in the context of random matrix theory. 
Indeed, the eigenvalues of the classical ensembles of random matrices can be mapped onto the positions of classical particles at equilibrium, interacting 
via a logarithmic potential, a system thus called the ``log-gas''. In general these particles are also submitted to an external potential (quadratic for the so called Gaussian ensembles) \cite{Mehtabook,Forrester}. For many random matrix ensembles explicit formulae have been obtained
for the variance of the linear statistics in mathematics~\cite{JohanssonLS,Pastur,BaiYao,Sosh,bourgade,JohanssonLS2,BaiWangZhou,Gus2005} and in physics~\cite{BZ1993,Beenaker1963,Vivo2014,MC2020}. 
For smooth linear statistics several examples have been studied, beyond the center of mass \cite{NM2009}. For instance, the case $f(x) = x(1-x)$ has been considered in the context of chaotic transport through a cavity where the positions $x_i$'s, with $0 \leq x_i \leq 1$, map onto the eigenvalues of an $N \times N$ Jacobi random matrix \cite{Beenaker1963,sommers,Khor,OK2008,VMB2008,VMB2010,DMTV2011}.
The case $f(x) = x^q$ has been considered in the context of the R\'enyi entropy in a random pure state of a bipartite system -- in this case $q>0$ is the R\'enyi index~\cite{Nadal1,Nadal2} and for other calculations of the entanglement entropy \cite{CLM2015,BP2021}. Yet another instance is the case $f(x) = 1/x$ for Wishart-Laguerre ensemble that has been studied in the context of the Wigner time-delay distribution \cite{TM2013}. Another interesting application
is to  choose $f(x)$ as the indicator function of some interval ${\cal I}$, in which case ${\cal L}_N$ corresponds to the number of particles inside ${\cal I}$.
The study of its fluctuations is generically called the {full counting statistics} (FCS) and usually includes the study of cumulants of order
higher than two.
The FCS has been widely studied in the one-dimensional log-gas in the context of random matrix theory \cite{Mehtabook,Forrester}, see e.g. \cite{MNSV2009,MNSV2011,MV2012,MMSV2014,MMSV2014PRL,FLD2020} and for applications to fermionic systems \cite{MMSV2016PRE,SmithCounting2021,SmithInteracting2021}. 
It is important to stress that the large $N$ limit of the linear statistics can be quite different depending on whether the
function $f(x)$ is smooth or an indicator function as for the FCS. For instance, for random matrix ensembles with Dyson index
$\beta=2$, Szeg\"o theorem type results apply in the first case~\cite{JohanssonLS}, while in the second case more complicated
Fisher-Hartwig type asymptotics are needed \cite{AbanovFH2011,DIK2011,Charlier2018,CC2020}.

Other long range interacting particle systems have been studied beyond the one dimensional log-gas, both in statistical 
physics \cite{Dauxois} and in mathematics \cite{lewin,Serfaty_lecturenotes}. 
One important example is the so-called (one component) ``Coulomb gas'' (CG), where $N$ particles with the same charge interact via the pairwise Coulomb potential in $d$ dimensions and in the presence of an external trap. Its thermodynamic properties were studied, e.g., in the context of plasma physics
\cite{Lenard,Baxter1963,AM1980,Choquard1981,JLM93,Janco95,forrester_el}.
More recently, there has been a renewed interest for Coulomb gases, due to their connections, in $d=2$, to non-Hermitian random matrices 
\cite{Mehtabook,Forrester},
as well as to the quantum Hall physics \cite{Cooper,Charles,oblak}, in particular in the context of non-interacting fermions in a rotating trap
\cite{Lacroix_rotating,Smith_rotating,Manas_rotating1,Manas_rotating2}. 
The FCS for the Coulomb gas {was studied in $d=2$ in \cite{allez,Lacroix_rotating,castillo,akemann1,akemann},
and in $d=1$ in \cite{dhar2017exact,Dhar2018,Flack22}.}

For smooth linear statistics, a CLT has also been established for the CG in various external potentials and in various dimensions~$d$ \cite{Forrester_91,RiderVirag2007,Ameur2011,Leble2018,BBNY,Flack2023,Serfaty2023}. These works yield quite general formulae for the first two cumulants of ${\cal L}_N$ to leading order at large $N$. Until recently there were no analogous results for the higher order cumulants, beyond the variance, of the linear statistics of the CG. In Ref. \cite{UsCoulomb}, these higher cumulants have been computed for the CG in any space dimension $d$ in the case of rotationally invariant external potential and linear statistics. The obtained formulae are surprisingly explicit in terms of the shape of the external potential and of the function $f$. 
These results were derived by computing the free energy of the CG to leading order at large $N$ in the presence of a tilted external potential. 
In this case the support of the equilibrium density of the gas is simply a sphere of radius $R$. 
Remarkably, it was found that, for any $d \geq 1$, while the variance depends on the values of the function $f$ 
inside the bulk of the gas, 
these higher order cumulants (of order $q \geq 3$) depend only
on the function $f$ at the edge of the gas. It is thus an outstanding problem to extend these results to 
more general interacting particle systems. 

The aim of this paper is to compute these higher order cumulants in the case of the one-dimensional Riesz gas where the $N$ particles interact
via a power law repulsive potential $\sim 1/|x_i-x_j|^k$, with $k > -2$. In addition the particles are confined by an external potential,
so that its mean (scaled) density has a finite support in the large $N$ limit. 
The statistical properties of the Riesz gas model have generated a lot of interest in recent years, both in the physics and mathematics literature~\cite{leble2017, leble2018, Agarwal_riesz, lewin, dereudre2020existence, kumar2020particles, boursier2021optimal, jit2021, boursier2022decay, santra2022, jit2022, lelotte2023phase, santra2023crossover, dereudre2023number, dandekar2023dynamical, hardin2017,Beenakker_riesz,Riesz_FCS}. Depending
on the value of $k$ the interaction is either long range $-2< k<1$, or short range for $k>1$. 
The class of Riesz gas includes three important well known models. 
 For instance, $k=2$ corresponds to the Calogero-Moser model ~\cite{calogero1971solution, calogero1975exactly, sutherland, moser, agarwalCM}, an integrable many-particle interacting system. On the other hand, the limit $k \to 0$ corresponds to the log-gas mentioned above \cite{Mehtabook},
 while the case $k=-1$ corresponds to the 1d CG also discussed above 
 ~\cite{lewin,Chafai22,flack2023out,rojas2018universal}. 
Interestingly, some fractional values of $k$ have been experimentally realized~\cite{joseph2011observation, zhang2017observation}. 
{The out of equilibrium dynamics of the Riesz gas has also been studied recently \cite{Mallick_riesz} using hydrodynamic methods, as well as 
its quantum generalisations \cite{Huse_riesz}.} 

In the present paper we extend the method used recently to compute the cumulants of the smooth linear statistics for the
Coulomb gas \cite{UsCoulomb} to the case of the Riesz gas. 
As we will see, for general $k$, it is technically much more challenging. 
In the case of the variance fairly explicit formulae
have been obtained recently \cite{Beenakker_riesz} for the long range case, for $-1<k<1$. 
In the short range case, a general formula for the variance of smooth linear statistics 
was obtained in \cite{Riesz_FCS}. Finally, some CLT results were also obtained for the Riesz gas on the circle
in the mathematics literature~\cite{boursier2021optimal,boursier2022decay}. At present however there are no results concerning the
higher order cumulants. The goal of this paper is to obtain such results, which even in the well studied case of the log-gas (i.e., $k=0$) do not
seem to be known. Since our method uses some large deviation techniques, we also have access to the large deviation form of
the full probability distribution of ${\cal L}_N$ at large $N$. 
In addition, in the short range case, we will also consider the FCS of the number of particles in a given interval (i.e., an 
example of non-smooth linear statistics), extending the results of Ref.~\cite{Riesz_FCS} beyond the variance.

The rest of the paper is organized as follows. In Subsection \ref{subsec:model}, we describe in detail the one-dimensional Riesz gas and in Subsection \ref{sec:density} we recall the main known results concerning the equilibrium density. In Section \ref{sec:mainresults}, we summarize our main 
results, both for the long range ($k<1$) and the short range ($k>1$) cases. In Section \ref{section:deriv}, we provide the derivation of these main results, treating separately the cases $k<1$ and $k>1$, before we conclude in Section \ref{sec:conclusion}. Several appendices contain of number of technical details.

\subsection{Model} \label{subsec:model}

In this paper we consider a confined Riesz gas at equilibrium. It consists of $N$ particles on the line in one dimension, interacting via a power law repulsive interaction $V(r) \sim 1/r^k$, where $r$ is the pairwise inter-particle distance, and confined by an external potential $U(x)$. We 
thus study the Gibbs distribution for the particle's positions $x_i$'s, $i=1,\dots,N$, namely 
\be  \label{eq:pdf}
{\cal P}(x_1,\dots,x_N) = \frac{1}{Z_{N,U}} e^{- \beta {\cal E}(\vec x)} 
\quad , \quad Z_{N,U} = \int_{\mathbb{R}^N} dx_1 \dots dx_N  \, e^{- \beta {\cal E}(\vec x)} 
\ee 
at an inverse temperature $\beta$, where the (potential) energy of the gas is 
\be \label{model} 
{\cal E}(\vec x) = N \sum_i U(x_i) + \frac{J {\rm sgn}(k)}{2} \sum_{i \neq j} \frac{1}{|x_i-x_j|^k} \;.
\ee 
Here $J > 0$ is the strength of the interaction, $k$ is the exponent of interaction and the sign function ${\rm sgn}(k)$ ensures that the interaction is repulsive. 
The case $k=-1$ corresponds to the (linear) Coulomb 
interaction in $d=1$. The limit $k \to 0$ corresponds to the log-gas. The 
case $k=2$ corresponds to the Calogero-Moser model.
The possible values of $k$ are such that the gas is confined and does not fly to infinity, e.g. for a
quadratic well $U(x)=x^2/2$ one must have $k>-2$. Here we focus on the case where $U(x)$ is an even function
of $x$, i.e., $U(x) = U(-x)$. 


An important quantity is the mean density defined as
\be 
\bar \rho_N(x) = \left\langle \frac{1}{N} \sum_{i=1}^N \delta(x-x_i) \right\rangle_{U}
\ee 
where $\langle \cdots \rangle_U$ denotes the average over the joint PDF in (\ref{eq:pdf}).
For later purpose it is convenient to indicate the external potential as a subscript. 
In \cite{Agarwal_riesz} the density of the gas in an harmonic trap was obtained.
It was found, and this is a generic feature, that there are two regimes: (i)
long range $k<1$, where the large scales dominate and dimensional analysis can be performed
by balancing the different terms in the energy, (ii) short range $k>1$, where the energy
becomes a local function of the density (it occurs when the sums $\sum_j 1/|x_j-x_i|$
for typically uniformly spaced particles are convergent at the microscopic scale). 

Note that in \eqref{model} we have chosen to multiply the external potential by a factor $N$. 
In the long range case $k<1$ this choice is convenient in order to study a large class of 
potentials $U(x)$, as it allows for the support
of the density to remain of order unity in the large $N$ limit.
The scaling with $N$ is thus {\it different} from the 
one used in \cite{Agarwal_riesz}
and instead, it is similar to the one used in our recent paper on the Coulomb gas in any space dimension \cite{UsCoulomb}. However, the two choices are easily related by a simple rescaling.
For a potential $U(x) \propto |x|^n$ ($n=2$ was studied there)
the connection with the conventions of Ref. \cite{Agarwal_riesz} is
then, denoting by tilde the same quantities in that paper
\be 
\tilde x = x N^\alpha \quad , \quad \tilde \beta \tilde {\cal E} = \beta {\cal E} 
\ee 
where $\alpha=1/(k+n)$ and $\tilde \beta \tilde {\cal E} \sim \tilde \beta N^{\frac{k+2 n}{k+n}}$
while here $\beta  {\cal E} \sim N^2$. To strictly match the Gibbs measures in both works we thus also need to
scale the temperature as $\beta = \tilde \beta N^{-k/(k+n)}$. Since the calculations
there and here are based on minimizing the energy, this is immaterial,
as long as $\tilde \beta \tilde {\cal E} = \beta {\cal E}   \gg N$, which is the scale of the entropy. 

\subsection{Equilibrium density in the large $N$ limit}\label{sec:density}

Let us recall how to obtain the equilibrium density in the large $N$ limit,
following \cite{Agarwal_riesz}.  
It is convenient to introduce the empirical density defined as 
\be 
\rho_{N}(x) =  \frac{1}{N} \sum_{i=1}^N \delta(x-x_i)  \;,
\ee 
where the ${x}_i$'s are distributed according to Eq. (\ref{eq:pdf}). 

\subsubsection{Long range case $k<1$} 

The energy in Eq. (\ref{model}) can be rewritten as a functional of the
density, which in the large $N$ limit takes the form, for $k<1$ 
\bea 
&& 
{\cal E}(\vec x) =  N^2 E[\rho_N] + O(N) \\ 
&& E[\rho] =  \frac{J {\rm sgn}(k)}{2}  \int_{\mathbb{R}} dx \int_{\mathbb{R}} dx'  
\frac{\rho(x) \rho(x') }{|x-x'|^k} +  \int_{\mathbb{R}} dx \rho(x) U(x) \;.
\eea 
As $N \to \infty$, $\rho_{N}(\bx)$ coincides with its average $\bar \rho_{N}(\bx)$,
and both converge to the equilibrium density $\rho_{\rm eq}(x)$ which is given by the minimizer of $E[\rho]$, 
under the additional constraint $\int_{\mathbb{R}} \rho(\bx) d\bx = 1$. 
To determine the minimizer one takes a functional derivative of $E[\rho]$ with respect to (w.r.t.) $\rho(\bx)$. This leads to
the following condition 
\bea \label{mini1}
 J {\rm sgn}(k) \int_{-\ell_0/2}^{\ell_0/2} dx'  \frac{1 }{|x-x'|^k} \rho_{\rm eq}(x')  +  U(x) + \lambda = 0 \;, \quad 
\eea
which is valid for any $x$ in the support $[-\ell_0/2, \ell_0/2]$ of $\rho_{\rm eq}(x)$. We assume here that
$U(x)$ is an even function of $x$ and such that the support is a single interval, with $\rho_{\rm eq}(\bx) = 0$ outside. Here  
$\lambda$ is a Lagrange multiplier enforcing the normalization of $\rho_{\rm eq}(x)$. Taking a derivative of Eq. (\ref{mini1}) with respect to $x$, one finds that the equilibrium density satisfies  
\be  \label{integral_eq}
J |k| \int_{-\ell_0/2}^{\ell_0/2} dx' \frac{{\rm sgn}(x'-x)}{|x-x'|^{k+1}} \rho_{\rm eq}(x') = - U'(x) \;,
\ee 
for $k \neq 0$. It is useful to introduce the parameter $p=k+1$. 
The solution of \eqref{integral_eq} is given in \cite{Agarwal_riesz} in the case of the
harmonic potential, and is based on the Sonin formula \cite{Sonin,stanley_riesz,Beenakker_riesz,Beenakker_stack}
which allows to treat more general cases. 
Since the gas must be stable and we are considering the long range case, 
one must have $-n < k < 1$, and the range of values of $p$ is thus $-n+1<p<2$. 
For that range, denoting $z=x +\ell_0/2$ and $p=k+1$ the density reads
\bea \label{rhoeq0} 
&&  \rho_{\rm eq}(x=z-\ell_0/2) = A z^{p/2-1} \frac{d}{dz} \int_z^{\ell_0} dt t^{1-p} (t-z)^{p/2} 
 \frac{d}{dt} \int_0^t dz' (z')^{p/2} (t-z')^{p/2-1} \, U'(z'-\ell_0/2) \\
 && A = - \frac{1}{J |k|} 
 \frac{2 \sin(\pi p/2)}{\pi p B(p/2,p/2)} \;,
\eea 
where the size of the support $\ell_0$ is determined by the normalization condition $\int_{-\ell_0/2}^{\ell_0/2} \rho_{\rm eq}(x)=1$. 
Note that in the full solution there is an additional zero mode for the density $c_0 (z (\ell-z))^{p/2-1}$ but one can argue, see 
\cite{Agarwal_riesz}, that $c_0$ must be set to zero. This is the case when the support length $\ell_0$ is allowed to
vary (as is the case here).

\subsubsection{Short range case $k>1$}

The short range case $k>1$ can be treated separately. For this case it is more convenient to
use the conventions of \cite{Agarwal_riesz} 
and define the original model as
\be \label{modelSR} 
{\cal E}(\vec x) = \sum_i U(x_i) + \frac{J {\rm sgn}(k)}{2} \sum_{i \neq j} \frac{1}{|x_i-x_j|^k} \;.
\ee 
Here we will restrict to power law potentials
\be 
U(x)= u_n |x|^n \;.
\ee 
In the large $N$ limit it was shown in \cite{Agarwal_riesz} that the energy reads 
\bea \label{E_SR}
&& {\cal E}(\vec x) =  N \int u_n |x|^n \rho_N(x) dx + J \zeta(k) N^{k+1} \int \left[ \rho_N(x)\right]^{k+1} \, dx \;.
\eea
One now defines 
\be 
x = y N^{\alpha_k } \quad , \quad \alpha_k=k/(k+n) \quad , \quad \tilde \rho_N(y) dy = \rho_N(x) dx  \;.
\ee 
Upon rescaling the energy becomes 
\be 
{\cal E}(\vec x) = N^{\frac{(n+1)k+n}{k+n}} E[\tilde \rho_N]
\ee 
with 
\be \label{Erho} 
E[\rho] =  \int u_n |y|^n \rho(y) dy + J \zeta(k)  \int \left[ \rho(y)\right]^{k+1} \, dy  \;.
\ee 
One can now define the rescaled partition function over the density field 
\be \label{partitionSR}
\tilde Z_{N,u_n |y|^n} = \int {\cal D} \rho  \; e^{- \beta N^{\frac{(n+1)k+n}{k+n}} E[\rho]  } \; \delta\left(\int dy \rho(y) -1\right)
\ee 
where the delta-function ensures the normalization of the density $\rho(y)$. One defines now the
rescaled averaged density as
\be 
\bar \rho_N(y) = \left\langle \frac{1}{N} \sum_{i=1}^N \delta(y-y_i) \right\rangle_{u_n\, |y|^n} \;,
\ee 
where the average is taken w.r.t. the Gibbs weight associated to the partition function in \eqref{partitionSR}. 

In the large $N$ limit the partition sum is dominated by the minimum of $E[\rho]$
and 
$\bar \rho_N(y)$ converges to $\tilde \rho_{eq}(y)$ which is the minimizer of $E[\rho]$ defined
in \eqref{Erho}, upon adding a Lagrange multiplier $\mu ( \int dx \rho(y)-1) $ to take into account the normalization constraint.
The minimization of this energy functional leads to the following equation for the large $N$ equilibrium density,
which is supported over the interval $[-\ell_0/2,+\ell_0/2]$
\bea 
u_n |y|^n + J \zeta(k) (k+1) \rho_{eq}^k(y) = u_n \left( \frac{\ell_0}{2} \right)^n \;,  \label{sp_sr}
\eea
which is solved as
\be  \label{ak_SR}
\rho_{eq}(y) = a_k u_n^{1/k} \left( \left(\frac{\ell}{2} \right)^n - |y|^n \right)_+^{1/k} \quad , \quad a_k=(J \zeta(k) (k+1))^{-1/k} \;.
\ee 
The normalisation condition gives
\be 
1 = \int_{-\ell_0/2}^{\ell_0/2} dy \rho_{eq}(y) = 2 a_k  u_n^{1/k} \left(\frac{\ell_0}{2}\right)^{1+ \frac{n}{k}}  \int_0^1 dy (1-y^n)^{1/k} 
\ee 
which leads to
\be \label{ell0}
\ell_0 = 2 (u_n)^{-\frac{1}{k+n}} \left( (2 a_k)^{-1} {\frac{1}{k}+ \frac{1}{n} \choose \frac{1}{k} }\right)^{ \frac{k}{k+n} } \;.
\ee 
In the case of the harmonic potential $n=2$, $u_n=1/2$, one finds \cite{Agarwal_riesz} 
\bea \label{dens_kgeq1}
\rho_{eq}(y) = \frac{1}{\ell_0}  F_k \left( \frac{y}{\ell_0}\right) \;, \; F_k(z) = \frac{1}{B(\gamma_k+1,\gamma_k+1)} \left(\frac{1}{4}-z^2\right)^{\gamma_k}
\quad , \quad \ell_0 = \left( \frac{(2 J \zeta(k) (k+1))^{1/k} }{ B(1 + \frac{1}{k},1 + \frac{1}{k}) } \right)^{\frac{k}{k+2}}
\eea
where $\alpha_k = k/(k+2)$ and $\gamma_k = 1/k$.
\\

\section{Observables and main results} \label{sec:mainresults} 

\subsection{Observables}

In addition to the mean density of the trapped gas, we are interested in the linear statistics, i.e., the fluctuations of ${\cal L}_N$ defined as
\begin{align}
  {\cal L}_N = \sum_{i=1}^N f(x_i)\,,\label{eq:lin}
\end{align}
where $x_i$'s are the positions of the particles, and $f(x)$ is a sufficiently smooth function. This excludes a priori indicator functions which appear in the
counting statistics problem, (see below for a discussion in a special case). To study the distribution of the linear statistics in (\ref{eq:lin}), one defines its cumulant generating function CGF $\chi(s,N)$ as
\begin{align}
 \chi(s,N) =  \log \langle e^{- N s \, {\cal L}_N } \rangle_U = 
 \log \langle e^{- N s \sum_{i=1}^N f(x_i)}\rangle_U \,,\label{eq:chidef}
\end{align}
where we recall that the average $\langle \cdots \rangle_U$ is taken over the joint PDF in (\ref{eq:pdf}). 

We will compute the CGF in the limit of large $N$. From it we will first extract the 
cumulants of ${\cal L}_N$, denoted as $\langle {\cal L}_N^q \rangle_c$, which are obtained from an expansion in
the parameter $s$
\be \label{cum_gen}
\chi(s,N) = \sum_{q \geq 1} \frac{(-1)^q}{q!} N^q s^q \langle {\cal L}_N^q \rangle_c \;,
\ee 
where $s$ is sufficiently small to ensure the convergence of the series. Note that 
in this definition of the generating function, the Laplace parameter is scaled as $N s$ with $s = O(1)$ -- or 
with another positive power of $N$ in the short range case -- see below in Eq.~(\ref{def_chi_SR}) -- hence it
explores the regime of large deviations of ${\cal L}_N$. By taking $s \to 0$ we expect to retrieve the
cumulants of ${\cal L}_N$. A more rigorous calculation of the cumulants (but much more difficult) would require instead to
scale $s=O(1/N)$. We will assume everywhere in this work, as in our previous works \cite{UsCoulomb,Flack2023}, 
that this will lead to the same result. A proof of this assumption remains however an outstanding problem.

Next we will be interested in the full probability distribution of ${\cal L}_N$,
i.e. the PDF ${\cal P}({\cal L}_N)$. At large $N$ we expect that it takes the large deviation form
\be \label{PP}
{\cal P}({\cal L}_N) \sim e^{- \beta N^2 \Psi(\Lambda) } \quad , \quad \Lambda=\frac{1}{N} {\cal L}_N = \frac{1}{N} \sum_i f(x_i) \;.
\ee 
The rate function $\Psi(\Lambda)$ is related to the CGF by a Legendre transform.
Indeed, inserting this large deviation form (\ref{PP}) in the definition of the CDF in Eq. (\ref{eq:chidef}) and performing the change of variables ${\cal L}_N \to \Lambda$ one obtains
\bea \label{eq:saddle_point}
\chi(s,N) \approx \log \left( \int_0^\infty d\Lambda \; e^{-N^2 \left( \beta \Psi(\Lambda) + s\, \Lambda\right)}\right) \;.
\eea 
For large $N$ the integral over $\Lambda$ can be evaluated by the saddle point method, leading to the relation 
\be 
\lim_{N\to \infty}\, \frac{1}{N^2}\chi(s,N) = - \min_{\Lambda \in \mathbb{R^+}} ( \beta \Psi(\Lambda) + s \,\Lambda) \;.  \label{legendre00} 
\ee 
One method to obtain $\Psi(\Lambda)$ is thus to extract it from $\chi(s,N)$ by Legendre inversion of (\ref{legendre00}). 

\subsection{Main results in the long range case $k<1$} 

\subsubsection{Cumulants of the linear statistics}

We obtain an explicit formula for the large $N$ limit of the cumulants, in the
case of power law potentials and power law linear statistics, i.e. for 
\be 
U(x) = u_n |x|^n \quad , \quad f(x) = f_m |x|^m \;,
\ee
where $m,n>1$. As mentioned above, it is convenient to use the variable $p=k+1$.
Let us recall that we are considering here the range $-n < k < 1$, hence $-n+1<p<2$. 
Our result for the first cumulant reads
\be 
\langle {\cal L}_N \rangle \simeq N m f_m \frac{c_{n,m}}{c_n} \ell_0^m \;, \label{average}
\ee 
where, for clarity we used the notation $\langle \ldots \rangle$ instead of $\langle \ldots \rangle_{u_n x^n}$.
In this formula $\ell_0$ is the size of the support of the Riesz gas at equilibrium in the
potential $U(x)$. It is given by the (normalization) condition
\be  \label{condnorm} 
n u_n c_n \ell_0^{p+n-1} = 1 \;.
\ee 

We now give the expression for the higher cumulants in the case where $n$ and $m$ are both even integers,
for which the expressions are quite explicit. 
The variance is obtained as
 \bea \label{var1} 
&&   \langle {\cal L}_N^2 \rangle_c 
 \simeq   \frac{1}{\beta} (m f_m)^2 c_{m,m} \ell_0^{2 m+p-1} 
\frac{m}{m+p-1}  \;.
\eea
We note that the variance has the special property that it depends on the potential $u_n$ only through the size of the support $\ell_0$.
The higher order cumulants with $q \geq 2$ read at large $N$
 \bea \label{higherc} 
&&   \langle {\cal L}_N^q \rangle_c 
 \simeq \frac{1}{\beta^{q-1}} \frac{(m f_m)^q (n u_n)^{1-q}}{N^{q-2}}    \frac{c_{m,m}}{c_m} \left(\frac{c_m}{c_n}\right)^{q-1} \ell_0^{q(m-n)+n} 
\frac{m}{m+p-1}  
\left[ \left(  a(\lambda)  \frac{d}{d\lambda} \right)^{q-2} \lambda^{2 m - 1+p} \right]_{\lambda=1} \\
&& {\rm with} \quad a(\lambda) := \frac{\lambda^{m+p}}{m+p-1 - (m-n) \lambda^{n-1+p}} \;.
\eea
In this formula, the coefficients $c_n$ and $c_{n,m}$ are given by (we recall that here $-1<p<2$)
\be \label{formulacm} 
c_2=
\frac{ \sin
   \left(\frac{\pi  p}{2}\right)
   \Gamma
   \left(\frac{p}{2}+1\right)^2}{\pi J p | k|  \Gamma (p+2)} \quad , \quad 
   \frac{c_m}{c_2}= \frac{3 \times 5 \dots (m-1)}{2^{m-2} (3+p) (5+p) \dots (m-1+p)} 
= \frac{2^{3-m} \Gamma
   \left(\frac{m+1}{2}\right) \Gamma
   \left(\frac{p+3}{2}\right)}{\sqrt
   {\pi } \Gamma \left(\frac{1}{2}
   (m+p+1)\right)}
\ee 
and 
\be  \label{c22nm} 
c_{2,2} = \frac{\sin
   \left(\frac{\pi  p}{2}\right)
   \Gamma
   \left(\frac{p}{2}\right)^2}{32
   \pi  J |k| (p+1) (p+3) \Gamma (p)} \quad , \quad 
\frac{c_{n, m} }{c_{2,2}} = 
\frac{2^{-m-n+8} \Gamma
   \left(\frac{m+1}{2}\right) \Gamma
   \left(\frac{n+1}{2}\right) \Gamma
   \left(\frac{p+1}{2}\right) \Gamma
   \left(\frac{p+5}{2}\right)}{\pi 
   m (m+n+p-1) \Gamma
   \left(\frac{1}{2} (m+p+1)\right)
   \Gamma \left(\frac{1}{2}
   (n+p-1)\right)}
\ee 
where 
\be 
c_{2,2} = \frac{\sin
   \left(\frac{\pi  p}{2}\right)
   \Gamma
   \left(\frac{p}{2}\right)^2}{32
   \pi  J |k| (p+1) (p+3) \Gamma (p)} \;.
\ee 
\\

For general $n,m>1$ we have also derived a formula for the higher order cumulants, which is given in Eq. \eqref{cumul_gen}. 
It also involves the coefficients $c_n$ and $c_{n,m}$ which are defined in terms of some multiple integrals, which
for even integer reproduce the above formula.

We have checked that the formulae above agree with previously known results which are available in the literature in special cases.
More specifically:
\begin{itemize} 
\item
{\bf The Coulomb Gas limit $p=0$}. For $k=-1$, i.e. $p=0$, the cumulants have been computed in \cite{UsCoulomb} (in any space dimension). Let us recall
here the formula in space dimension $d=1$ for the second and third cumulants (where $f(x)$ is an even function of $x$):  
\begin{eqnarray} \label{secondcum}
&& \langle {\cal L}_N^2 \rangle_c \simeq \beta^{-1} \int_0^{R} dx  [f'(x)]^2  \quad , \quad  \; \langle {\cal L}_N^3 \rangle_c \simeq  \frac{1}{\beta^2 N} 
\frac{ f'\left(R\right)^3}{
   U''\left(R\right)}  \label{third_cumul} \;,
\end{eqnarray} 
where $R$ is the solution of $U'(R)=1$. To compare with the present results we identify $R=\ell_0/2$. 
Since one has 
\be 
c_n|_{p=0,J=1,|k|=1} = 2^{1-n} \quad , \quad c_{n,m}|_{p=0,J=1,|k|=1} = \frac{n-1}{2^{m+n-1} m (m+n-1)} \;, \label{cnCG} 
\ee
we see that the normalisation condition \eqref{condnorm} becomes indeed $U'(R)=U'(\ell_0/2)=u_n (\ell_0/2)^{n-1} = 1$. 
Our formula~\eqref{var1} gives the variance for even integer $m$ as
 \bea \label{var1.2} 
  \langle {\cal L}_N^2 \rangle_c 
 \simeq  \beta^{-1} (m f_m)^2  
 \frac{1}{(2m-1)} 
 (\ell_0/2) ^{2 m-1}  \;,
\eea
which is in agreement with \eqref{third_cumul} setting $f'(x)=m f_m x^{m-1}$ and $R=\ell_0/2$. Similarly one finds from \eqref{higherc} the third cumulant as
\be 
\langle {\cal L}_N^3 \rangle_c \simeq  \frac{1}{\beta^2 N}  \frac{(m f_m)^3 }{ n (n-1) u_n} (\ell_0/2)^{3 m-n-1} \;,
\ee 
which is again in agreement with \eqref{third_cumul}, using (\ref{condnorm}). Below we will check that this agreement extends to all
cumulants. In addition we discuss the general case $m,n>0$. 



\item
{\bf The log-gas limit $p=1$}.
The case of the log-gas is obtained in the limit $k \to 0$, with $J k = 1$ fixed. 
In that case, only the variance (and covariance) was known previously for arbitrary smooth function $f(x)$.
In Appendix \ref{app_log_gas} we show that our results for the variance are in agreement with those
obtained in Refs. \cite{Vivo2014,Beenakker_loggas,Lambert} (when specified to monomial $f(x)$). In addition
in the present paper we also obtain the formula for the higher cumulants of the log-gas. These formulae can be checked
in the special case $m=n$, using a simple rescaling method. This is
detailed in Appendix \ref{app_log_gas}. 

\item 
{\bf The variance for $0<p<2$}. For $-1<k<1$ the variance of the linear statistics 
has been recently obtained by a different method in Ref. \cite{Beenakker_riesz} for arbitrary smooth function $f(x)$.
While in the general case the formulas involve complicated integrals, for monomial $f(x)$ these formulae are explicit.  
The detailed comparison with our result for the variance is given in Appendix \ref{app:Beenakker}.

\end{itemize}



\subsubsection{Large deviation rate function for the PDF ${\cal P}({\cal L}_N)$}

Through Legendre inversion of $\chi(s,N)$, see Section \ref{section:LDF}, we obtain that the PDF ${\cal P}({\cal L}_N)$ takes the large deviation
form \eqref{PP}, with the following rate function, for general $m,n>1$ 
\be
\Psi(\Lambda) = (n u_n) \frac{c_{m,m} c_n}{c_m^2} \ell_0^n \, \tilde \Psi(\tilde \Lambda) 
\quad , \quad  \Lambda =   (m f_m) \frac{c_{m,m}}{c_m} \ell_0^m
 \, \tilde \Lambda \;, \label{psi_main}
\ee 
in terms of a rescaled rate function $\tilde \Psi(\tilde \Lambda)$
determined 
from the parametric system
\bea \label{systemPsi} 
&& \tilde \Lambda = 
 \lambda^m - (1- \frac{c_m c_{n m}}{c_n c_{m,m}} ) \, \lambda^{p+n-1+m} \;, \\
&& \tilde \Psi'(\tilde \Lambda)  = \lambda^{n-m} -  \lambda^{-(p+m-1)} \;. \nonumber 
\eea
The typical/average value of $\tilde \Lambda$ corresponds to $\lambda=1$ hence to 
\be 
\tilde \Lambda_{\rm av} = \frac{c_m c_{n m}}{c_n c_{m,m}} \;,
\ee 
which agrees with the result \eqref{average} for the first cumulant (i.e. the average) of ${\cal L}_N$. 
Let us make several remarks. 

\begin{enumerate}
   
\item One can derive a parametric representation of $\Psi(\Lambda)$
in terms of a polynomial of $\lambda$, obtained 
by integrating the second equation in \eqref{systemPsi} 
\be 
\tilde \Psi(\tilde \Lambda)  =
\int_1^\lambda (\lambda^{n-m} -  \lambda^{-(p+m-1)}) \frac{d \tilde \Lambda}{d \lambda} d\lambda \label{int_psi}
\ee 
which we do not display here. 

\item Let us recall that for $n,m$ even integers the first equation in 
\eqref{systemPsi} simplifies into
\be  \label{ltilde}
 \tilde \Lambda = 
 \lambda^m - \frac{m (m-n)}{(m+n+p-1)(m+p-1)}  \, \lambda^{p+n-1+m} \;.
\ee 
In that case the parametric representation of $\tilde \Psi''(\tilde \Lambda)=\frac{d \tilde \Psi'(\tilde \Lambda)}{d\tilde \Lambda}$, computed from
\eqref{systemPsi}, simplifies as one obtains $\tilde \Psi''(\tilde \Lambda)=\frac{m+p-1}{m} \lambda^{-(2 m+p-1)}$, showing
that $\tilde \Psi$ is a convex function (in the region where \eqref{systemPsi} holds).


\item One sees that if $m>n$, $\tilde \Lambda$ as a function of $\lambda$ is non-monotonous.
It has a maximum at $\lambda=\lambda_c$ such that $\frac{d \tilde \Lambda}{d \lambda}=0$, which is given by
\be  \label{lambdac}
\lambda_c^{p+n-1} = \frac{m}{(p+n+m-1) (1- \frac{c_m c_{n m}}{c_n c_{m,m}} )} =_{n,m \, \text{even}} \frac{m+p-1}{m-n} \;,
\ee 
and simplifies for $n,m$ even integers.
Hence the interval of variation of $\lambda$ is $ \lambda \in [0, \lambda_c]$,
where $\tilde \Lambda$ reaches a maximal value $\tilde \Lambda_{\max}$, which in the case $n,m$ even integers with $m>n$
reads
\be  \label{location}
\tilde \Lambda_{\max}  = \lambda_c^m \frac{n+p-1}{m+n+p-1} = \frac{n+p-1}{m+n+p-1} \left( \frac{m+p-1}{m-n}  \right)^{\frac{m}{p+n-1}} \;.
\ee 
This indicates the presence of a phase transition for the rate function $\Psi(\Lambda)$, hence the above expressions
are valid only for $\tilde \Lambda < \tilde \Lambda_{\max}$. For the case $m,n$ integer, one sees that the derivative of the right hand side of the second
equation in \eqref{systemPsi} also vanishes at
$\lambda_c$, leading to a finite $\tilde \Psi''(\Lambda)$ at the transition.
As we show in 
Appendix \ref{app:psilog}, in the case of the log-gas $k=0$, i.e., $p=1$,
our formulae above precisely recover the results obtained in \cite{Valov} where it was shown that this transition corresponds to an ``evaporation
transition'' (with a fractional order between $2$ and $3$). To describe this transition in more details, one could generalize the approach of \cite{Valov} and compute the
rate function $\Psi(\Lambda)$ for the PDF ${\cal P}({\cal L}_N)$ using the variational formula
\be 
\Psi(\Lambda) = \min_{\rho(x) } \bigg[ 
\int dx U(x) \rho(x) + \frac{J {\rm sgn}(k)}{2} \int dx dx'   \frac{\rho(x) \rho(x')}{|x-x'|^k} 
- \Sigma \left( \int dx f(x) \rho(x) - \Lambda\right) - \mu \left(\int dx  \rho(x) - 1\right) \bigg] 
\ee 
where $\Sigma$ and $\mu$ are Legendre parameters. As in the case of the log-gas, one expects a phase diagram with 
several phases. The evaporation transition then corresponds to 
the rightmost or leftmost particle separating from the rest of the gas and dominating the statistics of ${\cal L}_N$. 
Similar transitions often occur in other contexts of ``pulled log-gases'' in random matrix theory \cite{MV09,MS14}. 
We will not perform in detail this analysis here, however we note that the present calculation
allows to locate this transition for the Riesz gas, see Eq. \eqref{location}. 
Note finally that there is another transition in \cite{Valov} for $m<2$ where the density at zero becomes negative
and leads to two supports. We will not attempt to discuss that transition in the context of the Riesz gas. 

\item In the case of the Coulomb gas $p=0$, the PDF large deviation rate function $\Psi(\Lambda)$ 
was computed in \cite{UsCoulomb} (see Appendix B there). We have checked, see Appendix \ref{app:psilog},
that setting $p=0$ in our present formulae \eqref{systemPsi} and \eqref{psi_main}, reproduce correctly the results obtained there. In addition
we give the location of the evaporation transition for the Coulomb gas, see Appendix \ref{app:psilog}.


\end{enumerate}

\subsection{Main results in the short range case $k>1$}

In the case of a short range interaction, $k>1$, as detailed above we use a different scaling such that
the support of the density remains of order unity in the large $N$ limit in the variable $y= N^{-\frac{k}{k+n}} x$.
We will thus consider here the linear statistics which is now defined as
\be  \label{def_LN_y}
\tilde {\cal L}_N : = \sum_i f(y_i) = \sum_i f\left( \frac{x_i}{N^{\frac{k}{k+n}}}\right) \;,
\ee 
where we restrict ourselves to functions $f(y)$ which are smooth and even in $y$.
The generating function is now defined as
\be \label{def_chi_SR}
\tilde \chi(v,N) =  \log \langle e^{- v  N^{\frac{n k}{k+n}} \, \tilde {\cal L}_N } \rangle_{u_n |y|^n} 
= \sum_{q \geq 1} \frac{(-1)^q}{q!} N^{\frac{q n k}{k+n}} v^q \langle \tilde {\cal L}_N^q \rangle_c \;,
\ee 
where, as above, the expectation value $\langle \ldots \rangle_{u_n |y|^n}$ is over the partition sum \eqref{partitionSR},
and $\langle \tilde {\cal L}_N^q \rangle_c$ denotes the cumulant of order $q$ of $\tilde {\cal L}_N$.

\subsubsection{Cumulants of the linear statistics}

We now detail our results for the cumulants. We find the scaling with $N$ for integer $q \geq 1$
\be 
\langle \tilde {\cal L}_N^q \rangle_c = O \left( N^{1 - \frac{(q-1) n k}{k+n} } \right) \;.
\ee 
The first moment is given, for large $N$, by
\bea  \label{first_sr}
&& \langle \tilde {\cal L}_N \rangle  
\simeq N 
2^{- \frac{n}{k}} a_k u_n^{1/k} \ell_0^{ \frac{n+k}{k}} \int_{0}^{1} dz f\left(\frac{\ell_0}{2} z\right) \, \left(  1 - 
 z^n \right)^{\frac{1}{k}}  \;,
\eea  
where $a_k$ is given in Eq. (\ref{ak_SR}) and $\ell_0$ in Eq. (\ref{ell0}). 

To display our result for the variance, we first define the integrals 
\be
I_p 
= 
\int_{0}^{1} dz f(\frac{\ell_0}{2} z)^p \,  \left(1-z^n \right)^{\frac{1}{k} - 1 } \;. \label{def_I_sum}
\ee 
The variance is then given at large $N$ by
\be
{\rm Var} \, \tilde {\cal L}_N  = \langle \tilde {\cal L}_N^2 \rangle_c \simeq \beta^{-1} N^{1 - \frac{n k}{k+n}}  \frac{a_k}{k} \left(\frac{u_n}{2^n}\right)^{\frac{1}{k}-1}  \ell_0^{1-n+ \frac{n}{k}}  \left( 
    I_2 - \frac{I_1^2}{I_0} \right)  \;. \label{var_sr}
\ee 
In the case of the harmonic potential $n=2$, this formula is in agreement with the recent result given in \cite{Riesz_FCS}. 
Note that this formula only involves the function $f(y)$ but not its derivatives. Hence it can be applied also
to the counting statistics problem where $f(x)$ is an indicator function (as was checked in \cite{Riesz_FCS}).
As we will see, this is not the case for the higher cumulants, which involve higher derivatives of $f(x)$,
as we now discuss. 

It is possible to obtain the higher cumulants for a generic function $f(y)$, which is even in $y$. The general procedure is detailed below.
However the expressions are very ``bulky'', hence we will display here only the third cumulant, which reads
\bea 
&& \langle \tilde {\cal L}_N^3 \rangle_c \simeq \beta^{-2} N^{1-\frac{2n k}{k+n}} a_k 2^{\left(2-\frac{1}{k}\right) n}
   l_0^{\left(\frac{1}{k}-2\right) n+1} u_n^{\frac{1}{k}-2} \times \int_0^1 dz \bigg[ 
  \frac{
   I_1-I_0 f\left(\frac{l_0 z}{2}\right)}{I_0^3 k^2
   \left(1-z^n\right)^{2 - \frac{1}{k}}} \nonumber \\
&& \times \bigg( \frac{1}{n} \left(1-z^n\right) \left(I_0
   f\left(\frac{l_0}{2}\right)-I_1\right) \left((k (n-1)-n)
   \left(I_0 f\left(\frac{l_0 z}{2}\right)-I_1\right)-I_0 k l_0 z
   f'\left(\frac{l_0 z}{2}\right)\right) \nonumber \\
   && +
  I_0 (k-1)
   \left(f\left(\frac{l_0}{2}\right)-f\left(\frac{l_0
   z}{2}\right)\right) \left(I_1 - I_0 f\left(\frac{l_0
   z}{2}\right) \right) \bigg) \bigg] \;. \label{third_cumul_sr}
\eea 
Note that apart from a complicated integral, it only involves the integrals $I_0$ and $I_1$ given in \eqref{def_I_sum}.




We now specialize to power law linear statistics of the form
\be 
f(y) = f_m |y|^m \;, {\rm with} \; m \geq 0 \;.
\ee 
In that case we obtain a general formula for all cumulants of order $q \geq 2$. It reads 
\bea  \label{formulacum} 
&&  \langle \tilde {\cal L}_N^q \rangle_c 
\simeq \beta^{1-q} (-1)^{q+1} N^{1-\frac{(q-1) n k}{k+n}} 
A_q \bigg( \big[  g(t)^{\frac{m-n}{1 + (n/k)}}  (1 + \frac{m-n}{1+ (n/k)} t \frac{ g'(t)}{ g(t)})  \big]^{-1} \,  \partial_{t}  \bigg)^{q-1} (g(t))^{-\frac{1+ (n/k) + m}{1 + (n/k)}} h(t) \bigg|_{t=0}
\eea  
where
\be  \label{def_Aq}
A_q =  ( 2 a_k)^{-\frac{ q (m- n) +n}{1+n/k}} 
 u_n^{\frac{k-q (k+m)}{k+n}} f_m^q \;,
\ee
while the functions $g(t)$ and $h(t)$ are defined as
\bea  \label{gseries}
&& g(t) = 
\int_0^1 dz  \left[ 1 - z^n + t \left(1 - z^m \right) \right]^{1/k} = \sum_{j \geq 0} \frac{g_j}{j!} t^j \\
&& {\rm with} \quad g_j = \frac{\Gamma(1+\frac{1}{k})}{\Gamma(1+\frac{1}{k} -j)} 
 \sum_{r=0}^j \frac{(-1)^r  \Gamma(1+j)}{\Gamma(1+r) \Gamma(1+j-r)} 
\frac{\Gamma(1+\frac{1}{k}-j) \Gamma(\frac{1+m r}{n}) }{n \Gamma(1+ \frac{1}{k} - j + \frac{1+m r}{n} ) } \;, \nonumber 
\eea  
and
\bea \label{hseries} 
&& h(t) = 
\int_0^1 dz \, z^m \left[ 1 - z^n + t \left(1 - z^m \right) \right]^{1/k} = \sum_{j \geq 0} \frac{h_j}{j!} t^j \\
&& {\rm with} \quad h_j = \frac{\Gamma(1+\frac{1}{k})}{\Gamma(1+\frac{1}{k} -j)} 
 \sum_{r=0}^j \frac{(-1)^r  \Gamma(1+j)}{\Gamma(1+r) \Gamma(1+j-r)} 
\frac{\Gamma(1+\frac{1}{k}-j) \Gamma(\frac{1+m (r+1)}{n}) }{n \Gamma(1+ \frac{1}{k} - j + \frac{1+m (r+1)}{n} ) }\;. \nonumber
\eea  
Expanding \eqref{formulacum} in terms of the derivatives at $t=0$ of $g(t)$ and $h(t)$, 
and replacing these derivatives  $g^{(j)}(0)=g_j$ and $h^{(j)}(0)=h_j$ by the expressions
given above for $g_j$ and $h_j$ one obtains the expression for the $q$-th cumulant. 
In Table~\ref{table_CM}, we display the lowest cumulants obtained for the Calogero-Moser model, which corresponds to 
$k=2$, $n=2$.


\begin{table}
\begin{eqnarray}
\begin{array}{|c||c|c|c|}
\hline
\quad & m = 2 & m= 4 & m = 6 \\ 
\hline 
\hline
\quad & \quad  & \quad  & \quad \\
q = 1 & \dfrac{1}{\pi } & \dfrac{2}{\pi ^2} & \dfrac{5}{\pi ^3}  \\
\quad & \quad  & \quad  & \quad \\
\hline
\quad & \quad  & \quad  & \quad \\
q=2 & -\dfrac{1}{2 \pi } & -\dfrac{17}{2 \pi ^3} &
   -\dfrac{131}{\pi ^5}  \\
\quad & \quad  & \quad  & \quad \\ 
\hline 
\quad & \quad  & \quad  & \quad \\
q=3 & \dfrac{3}{4 \pi } & \dfrac{225}{2 \pi ^4} &
   \dfrac{44235}{4 \pi ^7} \\
 \quad & \quad  & \quad  & \quad \\   
\hline 
\end{array}
\hspace*{0.5cm}
\begin{array}{|c||c|c|c|}
\hline
\quad & m = 1& m= 3 & m = 5 \\ 
\hline 
\hline
\quad & \quad  & \quad  & \quad \\
q = 1 & \dfrac{8}{3 \pi ^{3/2}} & \dfrac{64}{15 \pi ^{5/2}} &
   \dfrac{1024}{105 \pi ^{7/2}}  \\
\quad & \quad  & \quad  & \quad \\
\hline
\quad & \quad  & \quad  & \quad \\
q=2 & \dfrac{4}{\pi ^2}-\frac{1}{2} & \dfrac{256}{9 \pi
   ^4}-\dfrac{5}{\pi ^2} & \dfrac{65536}{225 \pi
   ^6}-\dfrac{63}{\pi ^4} \\
\quad & \quad  & \quad  & \quad \\ 
\hline 
\quad & \quad  & \quad  & \quad \\
q=3 & \dfrac{12-\pi ^2}{2 \pi ^{5/2}} & -\dfrac{8 \left(39 \pi
   ^2-4480\right)}{105 \pi ^{11/2}} & \dfrac{128
   \left(4685824+430575 \pi ^2\right)}{32175 \pi
   ^{17/2}}\\
 \quad & \quad  & \quad  & \quad \\   
\hline 
\end{array}
\nonumber
\end{eqnarray}
\caption{Values of the reduced $q$-th cumulant $W_{n=2,m,q}$ of the linear statistics
$\tilde {\cal L}_N = \sum_i f_m |y_i|^m$, defined via 
$\langle \tilde {\cal L}_N^q \rangle_c \approx
 (-1)^{q+1} N^{2-q} A_q W_{n=2,m,q} $, for various values of $m$ and $q$, for the Calogero-Moser model $k=2$ and $U(y)= u_2 y^2$, obtained from Eq.~(\ref{formulacum}).
Here $A_q =  ( 2 a_2)^{-\frac{ q (m- 2) +2}{1+2/k}} 
 u_2^{\frac{2-q (2+m)}{4}} f_m^q$ from Eq. (\ref{def_Aq}) and $a_2=(3 J \zeta(2))^{-1/2}$ from (\ref{ak_SR}).}\label{table_CM}
\end{table}


\subsubsection{Cumulants of the counting statistics}

We have also obtained the cumulants of the number of particles ${\cal N}_L$ inside the interval $y \in [-L/2,L/2]$
in rescaled coordinates.
This amounts to consider a linear statistics of the form ${\cal N}_L = \sum_i f(y_i)$ 
with $f(y)= \theta(\frac{L}{2}-y)\theta(y+\frac{L}{2})$. In principle one could try to use the above formula
for the cumulants upon inserting this function $f(y)$. As we have checked by a first principle
calculation of the cumulants presented in Section \ref{subsubsec:counting}, this gives the correct result 
for the second cumulant, which is a function of the ratio $r=L/\ell_0$, namely
\be \label{var_FCS_res}
\langle {\cal N}_L^2 \rangle_c  \simeq \frac{1}{\beta} N^{1 - \frac{n k}{k+n}}
\frac{2 a_k}{k} u_n^{\frac{1}{k}-1}  \left(\frac{\ell_0}{2}\right)^{1 + n(\frac{1}{k}-1) }  \tilde g_{\frac{1}{k}-1}(1)
\times (1 - h_1(r)) \, h_{1}(r) \quad, \quad {\rm where} \quad , \quad r = \frac{L}{\ell_0} \;,
\ee
where $a_k$ is given in Eq. (\ref{ak_SR}), $\ell_0$ in Eq. (\ref{ell0}), and 
we have defined the auxiliary functions
\be \label{def_gtilde}
\tilde g_{b}(r) = \int_0^r dz (1- z^n)^{b} = r \, _2F_1\left(-b,\frac{1}{n};1+\frac{1}{n};r^n\right)
\quad , \quad \tilde g_{b}(1) = \frac{\Gamma
   \left(1+b\right) \Gamma
   \left(1+\frac{1}{n}\right)}{\Gamma
   \left(\frac{1}{n}+1+b\right)} \quad , \quad   h_j(r) = \frac{\tilde g_{\frac{1}{k}-j}(r)}{\tilde g_{\frac{1}{k}-1}(1)}
\ee  
In the case $n=2$ this formula agrees with the result obtained in Eqs. (14) and (41) in \cite{Riesz_FCS} (with the identification
$I(r^2, \frac{1}{2} , \frac{1}{k})=h_1(r)$, $\ell_0 \to \ell_0/2$ and $u_n=1/2$). 
\begin{figure}[t]
\includegraphics[width = 0.9\linewidth]{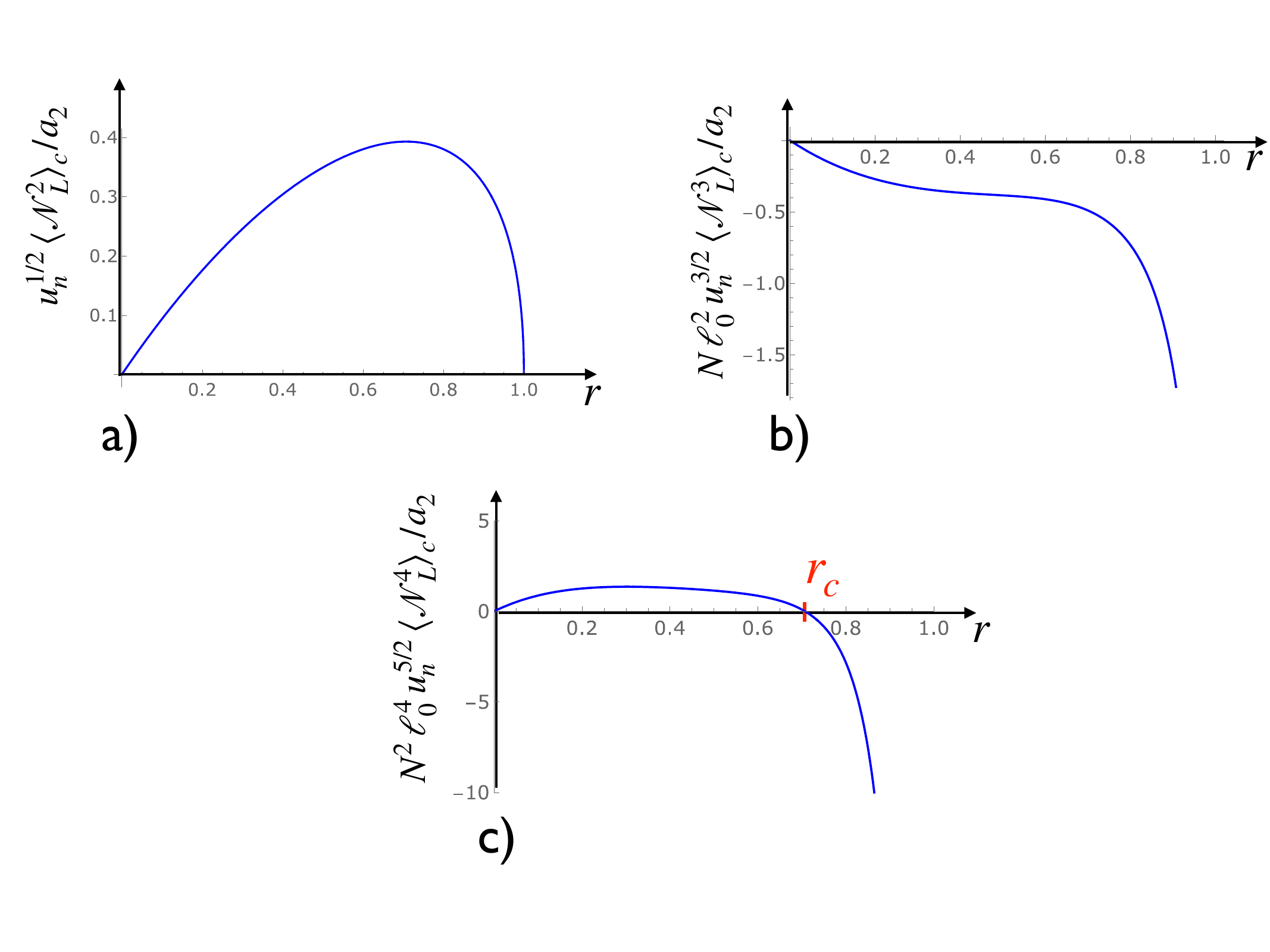}
\caption{Plot of the {second} (panel a), third (panel b) and fourth (panel c) scaled cumulants of ${\cal N_L}$ for the Calogero-Moser model, for inverse temperature $\beta = 1$, as given respectively in Eqs. \eqref{var_CM_intro}, \eqref{third_cumul_FCS} and \eqref{CMCS4}. While the second cumulant vanishes at $r=1$, the third and fourth cumulants diverge as $r \to 1$. {This feature is explained in the text}. In addition, we see that the fourth cumulant (panel c) changes sign from positive to negative as $r$ crosses the critical value $r_c = \sqrt{2}/2$.} \label{Fig_cumul}
\end{figure}
For the CM model $n=k=2$ the result is particularly simple 
\be \label{var_CM_intro}
\langle {\cal N}_L^2 \rangle_c \simeq 
\frac{1}{\beta} \frac{2 a_2}{\pi} {u_n^{-\frac{1}{2} }}  \sin ^{-1}(r) \cos ^{-1}(r) \;, 
\ee  
where we recall that $r=L/\ell_0$ and in that case $\ell_0^2 = 8/(\pi a_2 u_n^{1/2})$, with $a_k$ given in (\ref{ak_SR}) \cite{footnote1}.

In the present work we obtain the higher order cumulants of the counting statistics via a first principle calculation. For the third cumulant we obtain
\bea   \label{third_cumul_FCS}
&& \langle {\cal N}_L^3 \rangle_c \simeq - \frac{1}{\beta^2} N^{1-\frac{2 n k}{k+n}} \frac{2 a_k}{k} 
(\frac{\ell_0}{2} )^{1 + n (\frac{1}{k}-2) } { u_n^{\frac{1}{k} -2}}  \\
&& 
\times \left(h_1(r)-1\right) \tilde g_{\frac{1}{k}-1}(1) 
   \left(h_1(r){}^2 (\frac{1}{k} + \frac{1}{n} - 1) 
   +(\frac{1}{k}-1) 
   \left(h_1(r)-1\right) h_2(r)\right)- \frac{r}{n} h_1(r)
   \left(2 h_1(r)-1\right)
   \left(1-r^n\right)^{\frac{1}{k}-1} \;. \nn 
\eea   
For the CM model $n=k=2$ the result simplifies to
\be \langle {\cal N}_L^3 \rangle_c \simeq -
\frac{1}{N \beta^2} {u_n^{-\frac{3}{2} }} \frac{2 r a_2 \left(\pi ^2-12 \sin ^{-1}(r) \cos^{-1}(r)\right)}{\pi ^2 l_0^2 \sqrt{1-r^2}} \;.  \label{CMCS3} 
\ee
The general method to obtain the higher cumulants is presented in Section \ref{subsubsec:counting}. 
The formula for the fourth cumulant for general $n,k$ is rather ``bulky'', and given in 
Appendix \ref{App:FCS}. Here we display only the result for the CM model which reads 
\be 
\langle {\cal N}_L^4 \rangle_c \simeq
- \frac{{u_n^{-\frac{5}{2} }}}{N^2 \beta^3 } 
\frac{4 a_2 r \left(\pi -4 \sin ^{-1}(r)\right)
   \left(\pi  \left(6 \sqrt{1-r^2} \, r+\pi  \left(2
   r^2-3\right)\right)-8 \sin ^{-1}(r) \left(3
   \sqrt{1-r^2} \, r+\left(2 r^2-3\right) \cos
   ^{-1}(r)\right)\right)}{\pi ^3 l_0^4
   \left(1-r^2\right)^{3/2}} \label{CMCS4} 
\ee 
The second, third and fourth cumulants of ${\cal N}_L$ for the CM model are plotted in Fig. \ref{Fig_cumul}.
We see that the fourth cumulant changes sign at $r=r_c=\sqrt{2}/2$: it is positive for $r<r_c$ and negative for $r>r_c$.
More strikingly, while the second cumulant vanishes at $r=1$, the third and fourth diverge as $r \to 1$.
This divergence is discussed below.
\\

\noindent{\bf Remark}. As we noted above, inserting $f(y)= \theta(\frac{L}{2}-y)\theta(y+\frac{L}{2})$
in the general formula \eqref{var_sr} for the variance produces the correct result. If one
tries to do the same replacement in the general formula \eqref{third_cumul_sr} for the third cumulant,
one is faced with an ambiguity in evaluating the term $\propto f'(l_0 z/2)$. We have checked explicitly that using the rule
\be 
\left(\frac{I_1}{I_0} -f\left(\frac{\ell_0}{2} z\right)\right) f'\left(\frac{\ell_0}{2} z\right) = \frac{2}{\ell_0} \left(\frac{I_1}{I_0} -\frac{1}{2}\right) \delta(z-r)
\ee 
we can retrieve formula \eqref{third_cumul_FCS} for the third cumulant. For the higher cumulants
we have used the first principle calculation presented in Section \ref{subsubsec:counting} which does not contain any ambiguity. 
\\

\noindent{\bf Cumulants and probability distribution of ${\cal N}_L$ for $r=L/\ell_0$ close to unity}. 
We have obtained that for general $n,k$, the $q$-th cumulant behaves near $r=1$ as
\bea \label{cumul_r1}
&& \langle {\cal N}_L^q \rangle_c  \simeq \beta^{1-q} (-1)^{q} N^{1-\frac{(q-1) n k}{k+n}} 2 a_k 
\left(\frac{\ell_0}{2} \right)^{1 + n (\frac{1}{k}+1-q) } {u_n^{\frac{1}{k}+1-q}} 
n^{\frac{1}{k}+1-q} \frac{ \Gamma(1+ \frac{1}{k}) }{\Gamma(3-q+ \frac{1}{k})} (1-r)^{\frac{1}{k}+2 - q} \;.
\eea  
Hence, all cumulants of order $q \geq 3$ diverge as $r \to 1^-$. To understand this divergence
we recall that the full distribution ${\cal P}({\cal N}_L)$ of the number of particles ${\cal N}_L$ in $[-L,L]$ was studied in the case $n=2$ in \cite{Riesz_FCS}
where it was shown to take the large deviation form 
\be 
{\cal P}({\cal N}_L= c N ) \sim \exp \left( - \beta N^{1+ \frac{n k}{n+k}} \Phi(c,r) \right) \;,
\ee 
which we expect to be valid for all $n$. The rate function $\Phi(c,r)$ was given for $n=2$ in Eqs. (36)-(38) in \cite{Riesz_FCS}
and is recalled in Appendix \ref{app:prob} in the notations of the present paper. One can check explicitly,
see Appendix \ref{app:prob}, that for $r \to 1$ this rate function takes the following scaling form
\be  \label{def_PHI_txt}
\Phi(c,r) \simeq (1- r)^{\frac{1}{k} +2} \phi(\gamma) \quad , \quad \gamma = \frac{1-c}{(1-r)^{\frac{1}{k} +2}} \;.
\ee 
On the other hand by resuming our result for the cumulants in \eqref{cumul_r1} and performing a Legendre transform (see Appendix \ref{App:FCS}),
we obtain the same scaling form (\ref{def_PHI_txt}) for any $n,k$, with 
\bea  \label{phig} 
 && \phi(\gamma) =  \phi(0)  \left( 1 - \left(2 + \frac{1}{k}\right) \frac{\gamma}{\gamma^*} + \left(1 + \frac{1}{k}\right)  \left(\frac{\gamma}{\gamma^*}\right)^{\frac{1+2 k}{1+k} } \right) \\
&& \phi(0) = n u_n \left( \frac{\ell_0}{2}\right)^n \frac{k}{1+2k} \,\gamma^* 
\quad , \quad \gamma^* = {n^{1/k}} \frac{\Gamma(1 + \frac{1}{k} + \frac{1}{n})}{\Gamma(2 + \frac{1}{k}) \Gamma(1 + \frac{1}{n})  } \nonumber \;.
\eea 
For $n=2$ this formula agrees with the small $1-r$ expansion 
of the result of \cite{Riesz_FCS} performed in the Appendix \ref{app:prob}.
In~\eqref{phig} the value $\gamma=\gamma^*$ corresponds to the minimum of the scaled rate function with $\phi(\gamma^*)=0$
and gives the most probable, i.e., the average value 
\be 
\langle {\cal N}_L \rangle = N \left( 1 - \gamma^* (1-r)^{\frac{1}{k} + 1} + o((1-r)^{\frac{1}{k} + 1}) \right) 
\ee 
In addition our result for $\gamma=0$ gives the probability $P_L$ that the two intervals $(-\infty, - \frac{L}{2} ]  \cup 
[\frac{L}{2}, + \infty) $ are both empty of particles, when $r=L/\ell_0$ is close to unity. More precisely
\be \label{emptiness}
P_L \sim \exp \left( - \beta N^{1+ \frac{n k}{n+k}}  (1- r)^{\frac{1}{k} +2} \phi(0) \right) 
\ee 
where $\phi(0)$ is given in \eqref{phig}. We have checked (see Appendix \ref{app:prob}) that for $n=2$, $\frac{1}{2} \phi(0)$ agrees with
the result obtained in \cite{Riesz_FCS} for the rate function of the index in the limit $r=L/\ell_0$ is close to unity. {We have also checked that (\ref{emptiness}) agrees with the result obtained for the distribution of the position of the rightmost particle studied in Ref. \cite{jit2022} for the case $n=2$~\cite{footnote2}.

\section{Derivations} \label{section:deriv}

We sketch now the derivations of the results. We start with the long range case.

\subsection{Long range case $k<1$}

\subsubsection{Cumulants for power law linear statistics} 

Our aim now is to compute the generating function of the cumulants
\be \label{chi_deriv}
\chi(s,N) = \log \langle e^{- N s \sum_i f(x_i) } \rangle_U = \sum_{q \geq 1} 
\frac{(-1)^q}{q!}N^q s^q \langle {\cal L}_N^q \rangle_c \;.
\ee 
We will perform the calculation setting $\beta=1$ since the dependence on
$\beta$ of the $q$-th cumulant is simply $\propto \beta^{1-q}$, as can be seen by rescaling.
The generating function can be written as a ratio of two partition sums (as defined in Eq.~\eqref{eq:pdf}) 
\be  \label{ratio_deriv}
\chi(s,N) = \log \frac{Z_{N,U+ s f} }{Z_{N,U}}
\ee 
Taking a derivative w.r.t. $s$ one finds the exact relation
\be 
\partial_s \chi(s,N) = - N \Big \langle \sum_i f(x_i) \Big \rangle_{U+ s f} 
\ee 
In the large $N$ limit the right hand side (r.h.s.) converges to an average over the equilibrium density in the presence of the 
shifted potential $U+s f$ which we denote $\rho_{eq,s}(x)$, and one obtains, to leading order for large $N$, 
\be \label{ds} 
\partial_s \chi(s,N) \simeq  - N^2 \int dx f(x) \,  \rho_{eq,s}(x) \;.
\ee 
Hence one needs to find the equilibrium density in the shifted potential $U(x)+s f(x)$. 
In this paper we choose $f(x)$ to be an even function of $x$, so that the shifted 
potential is also even. To compute the cumulants we only need to consider $s$ vanishingly small,
hence we expect that the support of $\rho_{eq,s}(x)$ to be a single interval that we 
denote $[-\ell_s/2, \ell_s/2]$, which remains close to the equilibrium support $[- \ell_0/2,\ell_0/2]$ as $s \to 0$. 
Indeed in the finite region around $x=0$ including both supports the shifted potential
remains sufficiently confining at small $s$. 
\\

To determine $\rho_{eq,s}(x)$ we can thus use the 
formula \eqref{rhoeq0} by replacing there $U'(x)$ by $U'(x) + s f'(x)$ and $\ell_0$ and $\ell_s$.
From now on, as discussed in the introduction, we restrict to the case of monomials
\be 
U(x) = u_n |x|^n \quad , \quad f(x) = f_m |x|^m \label{u_f} \;,
\ee 
with $n,m>1$. Using Eq. \eqref{ds} together with the Sonin formula \eqref{rhoeq0} we obtain
\bea  
&& \partial_s \chi(s,N) \simeq  - N^2  f_m 
A  \int_{0}^{\ell_s} dz |z- \frac{\ell_s}{2}|^m
z^{p/2-1} \frac{d}{dz} \int_z^{\ell_s} dt \, t^{1-p} (t-z)^{p/2} \\
&& 
\times  \frac{d}{dt} \int_0^t dz' (z')^{p/2} (t-z')^{p/2-1} \, 
 \left( n u_n |z'- \frac{\ell_s}{2}|^{n-1} + m f_m s
 |z'- \frac{\ell_s}{2}|^{m-1} \right) 
 {\rm sgn}(z'- \frac{\ell_s}{2})  \;, \nonumber 
 \eea  
with
\be \label{def_A}
A = - \frac{1}{J |k|} \frac{2 \sin(\pi p/2)}{\pi p B(p/2,p/2)} \;,
\ee 
where $B(a,b) = \Gamma(a) \Gamma(b)/\Gamma(a+b)$ is the beta function. 
 Upon rescaling in that formula, $z' \to \ell_s z' $, $t \to \ell_s t$
 and $z \to \ell_s z $ we obtain
\be \label{tocompare} 
 - \frac{1}{N^2} \partial_s \chi(s,N)  \simeq m f_m \left( n u_n c_{n,m} \ell_s^{p+n-1+m} + s \, m f_m \, c_{m,m} \ell_s^{p+2 m-1}\right) \;,
\ee 
where the coefficients $c_{n,m}$ are defined by the rescaled integral formula 
\bea \label{definitioncnm}
&& c_n = A \int_0^1 dz z^{p/2-1} \frac{d}{dz} \int_z^1 dt t^{1-p} (t-z)^{p/2} 
 \frac{d}{dt} \int_0^t dz' (z')^{p/2} (t-z')^{p/2-1}  |z'-1/2|^{n-1} {\rm sgn}(z'-1/2) \;, \\
&& c_{n,m} = 
A \int_0^1 dz \frac{|z-1/2|^{m}}{m} z^{p/2-1} \frac{d}{dz} \int_z^1 dt t^{1-p} (t-z)^{p/2} 
 \frac{d}{dt} \int_0^t dz' (z')^{p/2} (t-z')^{p/2-1} |z'-1/2|^{n-1} {\rm sgn}(z'-1/2) \;. \nn
\eea
In addition we must impose the normalisation condition
\be \label{norm_rhos}
\int_{-\ell_s/2}^{\ell_s/2} dx \rho_{eq,s}(x) = 1 \;,
\ee 
which upon using \eqref{rhoeq0} 
under the same rescaling, gives the condition
\be \label{norm2} 
n u_n c_n \ell_s^{p+n-1} + s m f_m c_m \ell_s^{p+m-1} = 1 \;,
\ee 
which eventually determines $\ell_s$ as a function of $s$. 

The goal is in principle to eliminate $\ell_s$ from these equations
and compute the higher derivatives of $\partial_s \chi(s,N)$
w.r.t. $s$ at $s=0$ to obtain the cumulants. Since the equations
are algebraic and their solutions are not simple to express, we
choose a different route and instead eliminate $s$ from the
second equation \eqref{norm2} to express the first one only as a function of $\ell_s$.
This gives $s=g(\ell_s)$, which upon substituting into \eqref{tocompare} leads to the parametric system
\bea 
&& g(\ell_s) = \frac{1}{m f_m c_m} \ell_s^{-(p+m-1)} - \frac{n u_n c_n}{m f_m c_m} \ell_s^{n-m}  \;, \label{def_g1}\\
&& - \frac{1}{N^2} \partial_s \chi(s,N)  \simeq F(\ell_s) 
= m f_m \left( \frac{c_{m,m}}{c_m} \ell_s^m + n u_n \left(c_{n,m}- \frac{c_n c_{m,m}}{c_m} \right) \ell_s^{p+n-1+m} \right) \label{chiprime} \;.
\eea 
It is useful to rescale $\ell_s$ by $\ell_0$. Setting $s=0$ in \eqref{norm2} we obtain 
\eqref{condnorm}, i.e. $n u_n c_n \ell_0^{p+n-1} = 1$. Upon rescaling, 
the above system can be rewritten as
\bea 
&& g(\ell_s) = \frac{n u_n c_n}{m f_m c_m \ell_0^{m-n}} \tilde g\left(\frac{\ell_s}{\ell_0}\right) \quad , \quad \tilde g(\lambda)= \lambda^{-(p+m-1)}- \lambda^{n-m} \label{def_g2} \\
&& F(\ell_s)= \frac{m f_m c_{m,m}}{c_m} \ell_0^m
\tilde F\left(\frac{\ell_s}{\ell_0}\right) \quad , \quad 
\tilde F(\lambda) =
 \lambda^m - \left(1- \frac{c_m c_{n m}}{c_n c_{m,m}} \right) \, \lambda^{p+n-1+m} \;.  \label{Flambda} 
\eea 

We can now obtain an expression for the cumulants. We use that
\be 
\partial_s =  \frac{1}{g'(\ell_s) } \partial_{\ell_s} 
\ee 
and one finds for integer $q \geq 1$ 
\be
\langle {\cal L}_N^q \rangle_c =  \frac{1}{N^q} (- \partial_s)^q \chi(s,N)|_{s=0}
\approx \frac{1}{N^{q-2}} (- \partial_s)^{q-1} F(\ell_s)|_{s=0} 
= \frac{1}{N^{q-2}}  \left(\frac{-1}{g'(\ell)} \frac{d}{d\ell} \right)^{q-1} F(\ell)|_{\ell=\ell_0} \;.
\ee 
This leads to our main result for the cumulants in the case $m,n>1$ (restoring the $\beta$-dependence)
\be 
 \langle {\cal L}_N^q \rangle_c \approx 
 \label{cumul_gen}
 \frac{1}{\beta^{q-1}\,N^{q-2}}  (m f_m)^q (n u_n)^{1-q}  \frac{c_{m,m}}{c_m} \left(\frac{c_m}{c_n}\right)^{q-1} \ell_0^{q(m-n)+n} 
\left( a(\lambda)  \frac{d}{d\lambda} \right)^{q-1} \tilde F(\lambda)|_{\lambda=1} \;, 
\ee
where 
\be 
a(\lambda) = \frac{-1}{\tilde g'(\lambda)} = \frac{\lambda^{m+p}}{m+p-1 - (m-n) \lambda^{n-1+p}} \;,  \label{alambda} 
\ee 
and $\tilde F(\lambda)$ is given in \eqref{Flambda}. We recall that the coefficients $c_n$ and $c_{n,m}$
are defined in \eqref{definitioncnm}. 

The coefficients $c_n$ and $c_{n,m}$ can be explicitly evaluated for even integer $m,n$. 
The formula is given in the previous section in \eqref{c22nm}. A derivation of 
alternate but equivalent expressions are given in Appendix \ref{app:cm} -- see Eq. (\ref{cm_exact.1}) -- 
and in Appendix \ref{app:cnm}, see Eq. (\ref{cnm.1}).

In the case where $n,m$ are even integers, an additional simplification occurs. The following identity
is found to hold
\be \label{miracle} 
 1- \frac{c_m c_{n, m}}{c_n c_{m,m}} = \frac{m (m-n)}{(m+n+p-1)(m+p-1)} \;.
\ee 
It implies the remarkable relation
\be \label{simplegen} 
a(\lambda)  \frac{d}{d\lambda} F(\lambda)= \frac{m}{m+p-1}  \lambda^{2 m - 1+p}
\ee
and allows to rewrite the general formula for the cumulants \eqref{cumul_gen} in the more compact form
given in the previous Section, see Eqs. \eqref{var1} and \eqref{higherc}.

\subsubsection{Coulomb gas limit $k=0$} 

We can compare our formula for the cumulants \eqref{cumul_gen}
with the one obtained in \cite{UsCoulomb} for the case of the Coulomb gas $p=0$ and $J=1$.
There the following formula was obtained for the cumulants $q \geq 3$, with $r=|x|$ (specializing to $d=1$)
\be 
\langle {\cal L}_N^q \rangle_c \simeq \frac{1}{\beta^{q-1} N^{q-2}} \, (A(r) \partial_r)^{q-3}  \left( A(r)  f'(r)^2  \right)|_{r=R} 
\quad , \quad U'(R)  = 1 \;,  \label{CGcum}
\ee 
where $R=\ell_0/2$ and the function $A(r)$ is given by
 \be 
A(r) = 
\frac{ f'\left(r\right){}^2}{
   f''\left(r\right) \left(1 -
   U'\left(r\right)\right)+f'\left(r\right)
    U''\left(r\right)} \;.
   \ee 
Inserting $U(r)=u_n r^n$ and $f(r)=f_m r^m$ and 
setting $r=R \lambda$ one obtains
\be 
A(r) = m f_m R^m a(\lambda) \quad , \quad a(\lambda)=\frac{\lambda^m}{ m-1 + (n-m) \lambda^{n-1}} 
\ee 
where $a(\lambda)$ coincides with the function given in \eqref{alambda} for $p=0$. Inserting into \eqref{CGcum}
we find that the result in \cite{UsCoulomb} can be rewritten as
\be \label{cumCGgen} 
\langle {\cal L}_N^q \rangle_c \simeq \frac{1}{\beta^{q-1} N^{q-2}} \, (m f_m)^q R^{(m-1) q + 1} (a(\lambda) \partial_\lambda)^{q-3} 
(a (\lambda)\lambda^{2 m-2}) \;.
\ee

To compare with the above formula \eqref{cumul_gen}, we first note that the 
normalization condition $U'(R)=1$ is equivalent to the condition $n u_n R^{n-1}=1$.
Comparing with the condition \eqref{condnorm}, using $R=\ell_0/2$, we see that it implies
$c_n=2^{1-n}$ for $p=0$. Next we note the equality (using the normalization condition \eqref{condnorm} and again $R=\ell_0/2$)
\be 
(n u_n)^{1-q}   \ell_0^{q(m-n)+n} = 2^{q(m-n) + n} R^{q(m-1)+1} \;. \label{condnorm2}
\ee 
Furthermore for the formula \eqref{cumCGgen} and our formula obtained here \eqref{cumul_gen}
to agree for all $q \geq 3$, we see that one needs that the following terms must be
proportional for all $\lambda$
\be 
(a(\lambda) \partial_\lambda)^{2} \tilde F(\lambda) = B_m a(\lambda) \lambda^{2m-2} \label{constr} \;,
\ee 
where $B_m$ must be independent of $\lambda$. This is a rather strong constraint and one can check, remarkably
that it can be satisfied, and 
that it is equivalent to the two identities
\be
1- \frac{c_m c_{n, m}}{c_n c_{m,m}} = \frac{m (m-n)}{(m+n-1)(m-1)} \quad , \quad B_m= \frac{ m (2m-1)}{m-1} \;.
\ee 
Using \eqref{condnorm2} and \eqref{constr} and $c_n=2^{1-n}$ we see that the two formulae \eqref{cumCGgen} and \eqref{cumul_gen} are equivalent if and only if
in addition, one has
\be 
2^m \frac{c_{mm}}{c_m} \frac{ m (2m-1)}{m-1}  =1 \;.
\ee 
These conditions are compatible and, interestingly, they completely determine the coefficients $c_{n,m}$ and $c_n$ 
in the limit $p \to 0$, namely (with $J=1$)
\bea \label{cnm_p0}
c_n = 2^{1-n} \quad, \quad c_{n,m} = \frac{n-1}{2^{m+n-1} m (m+n-1)} \;.
\eea 
Until now all the above statements are valid for any $n,m$. In the case of even integer $n,m$ since we have 
explicit formula for these coefficients for general $p$ -- see Eqs. \eqref{formulacm} and \eqref{c22nm} -- one easily checks that setting $p \to 0$ in these formula all the conditions stated above are obeyed. This shows that the
results of the two works are in agreement. 

Since we know that, for $p=0$, the relation \eqref{miracle} is in fact correct for any $n,m>1$
it is natural to ask whether it may also hold for all $m,n$ beyond $p=0$.


\subsubsection{Large deviation rate function for the full probability distribution ${\cal P}({\cal L}_N)$} \label{section:LDF}

Let us recall that the full probability distribution of ${\cal L}_N$ takes the large deviation form (setting here $\beta=1$ as above)
\be \label{PP_text}
{\cal P}({\cal L}_N) \sim e^{- N^2 \Psi(\Lambda) } \quad , \quad \Lambda=\frac{1}{N} {\cal L}_N = \frac{1}{N} \sum_i f(x_i) \;,
\ee 
where the rate function $\Psi(\Lambda)$ is related to the generating function $\chi(s,N)$ by the Legendre transform formula
\be 
\lim_{N\to \infty}\, \frac{1}{N^2}\chi(s,N) = - \min_{\Lambda \in \mathbb{R^+}} ( \Psi(\Lambda) + s \,\Lambda) \;.  \label{legendre00text} 
\ee 
Let us now invert \eqref{legendre00text} to obtain $\Psi(\Lambda)$. A convenient way is to use the
parametric representation for $\chi(s,N)$ given in Eqs. \eqref{def_g1}-\eqref{chiprime}. Taking derivatives we obtain the system
\bea  
&& - \frac{1}{N^2} \partial_s \chi(s,N)  = \Lambda = F(\ell_s) \\
&& \Psi'(\Lambda) = - s = - g(\ell_s) 
\eea  
which by elimination of $\ell_s$ gives $\Psi'(\Lambda)$ as a function of $\Lambda$. 
For general $m,n>1$ we can now use the scaling form 
\eqref{def_g2} and obtain the result given in \eqref{int_psi} above.
The special case of the log-gas $p=1$ is discussed in the Appendix \ref{app:psilog}.

\subsection{Short range case $k>1$}

\subsubsection{General even linear statistics $f(y)$}

Let us now study the cumulant generating function for the short range case $k>1$. We recall
that we have rescaled the coordinates, writing $x = N^{\alpha_k} y$ with $\alpha_k=k/(k+n)$
so that the support of the density remains of order unity in the large $N$ limit in the variable $y$.
The linear statistics of interest is thus defined as 
\be  \label{def_LN_y2}
\tilde {\cal L}_N : = \sum_i f(y_i) = \sum_i f\left( \frac{x_i}{N^{\frac{k}{k+2}}}\right) \;,
\ee 
where we restrict ourselves to functions $f(y)$ which are smooth and even in $y$.
As for the long range case, since the dependence on $\beta$ of the $q$-th cumulant is simply $\propto \beta^{1-q}$
we will set $\beta=1$ in the calculation (and restore it at the end).
The generating function is defined as
\bea  
\tilde \chi(v,N) &=&   
 \log \langle e^{- v  N^{\frac{n k}{k+n}} \, \tilde {\cal L}_N } \rangle_{u_n |y|^n} 
= \sum_{q \geq 1} \frac{(-1)^q}{q!} N^{\frac{q n k}{k+n}} v^q \langle \tilde {\cal L}_N^q \rangle_{c} \label{sumsum} 
\\ &\simeq& \log \frac{\tilde Z_{N,u_n |y|^n+ v f(y)}}{\tilde Z_{N,u_n |y|^n}} \,,\label{eq:tildechidef}
\eea  
and can be rewritten in the large $N$ limit in terms of a ratio of partition sums computed with the shifted potential $|y|^n + v f(y)$ and the original potential $|y|^n$, respectively.
The expectation value $\langle \ldots \rangle_{u_n |y|^n}$ is taken over the partition sum which is defined in \eqref{partitionSR}. 
The expression $\langle \tilde {\cal L}_N^q \rangle_{c}$ denotes the $q$-th cumulant of $\tilde {\cal L}_N$, and
we will omit below the subscript $u_n |y|^n$.
Taking a derivative w.r.t $v$, one finds the relation
\bea 
&& \partial_v \tilde \chi(v,N)  
\simeq  - N^{\frac{(n+1)k+n}{k+n}} \,  \int dy f(y) \, \rho_{eq,v}(y) \;,
\eea 
where $ \rho_{eq,v}(y)$ is the equilibrium density in the variable $y$ in the shifted potential.
Since $f(y)$ is even the support is still a symmetric interval $[-\ell_v/2 , \ell_v/2]$.
Using Eq. \eqref{sp_sr} substituting the potential $u_n |y|^n$ by the shifted potential $u_n |y|^n + v f(y)$, we obtain that $ \rho_{eq,v}(y)$ must obey the relation
\bea 
J \zeta(k) (k+1) \left[\rho_{eq,v}(y) \right]^k = u_n \left(\frac{\ell_v}{2}\right)^n + v \frac{f(\frac{\ell_v}{2})}{\beta}  - \left( 
u_n |y|^n + v \frac{f(y)}{\beta} \right ) \;,
\eea 
where $\ell_v$ is determined by the normalization condition
\be 
\int_{-\ell_v/2}^{\ell_v/2} dy  \rho_{eq,v}(y) = 1 \;.
\ee

The cumulant generating function is thus determined by the solution of the following parametric system
\bea \label{system}
&& - N^{-\frac{(n+1)k+n}{k+n}}  \partial_v \tilde \chi(v,N) =  
a_k \int_{-\ell_v/2}^{\ell_v/2} dy f(y) \, \bigg[ g(v) 
 - \left( 
u_n |y|^n + v f(y) \right ) 
\bigg]^{\frac{1}{k}} \\
&& 1 = 
a_k  \int_{-\ell_v/2}^{\ell_v/2} dy \bigg[ g(v)  - \left( 
u_n |y|^n + v f(y) \right ) 
\bigg]^{\frac{1}{k}} \quad , \quad g(v) := u_n \left(\frac{\ell_v}{2}\right)^n + v f\left(\frac{\ell_v}{2}\right) \;, \nonumber 
\eea 
where we denote $a_k = \frac{1}{[J \zeta(k) (k+1)]^{1/k}}$. Taking $v=0$ in this system, rescaling
$y =\ell_0 z/2$ and using the symmetry $z \to -z$,
the first equation in \eqref{system} reproduces [using \eqref{sumsum}] the formula for the first moment given above in \eqref{first_sr}.
It depends only on $\ell_0$ which is itself determined by the second equation in \eqref{system} for $v=0$
\be 
1 = 
2 a_k u_n^{1/k} \left(\frac{\ell_0}{2}\right)^{1+ \frac{n}{k}} \int_{0}^{1} dz ( 1  - |z|^n )^{\frac{1}{k}} \;.
\ee
which leads to \eqref{ell0}.

Let us now compute the variance. To this aim let us define
\be \label{def_Jp}
J_p(v) = \frac{a_k}{k} \int_{-\ell_v/2}^{\ell_v/2} dy f(y)^p \bigg[ g(v)  - \left( 
u_n |y|^n + v f(y) \right ) 
\bigg]^{\frac{1}{k} - 1 } \;.
\ee 
Taking the derivative of the second equation in \eqref{system} w.r.t. $v$ one sees that the boundary term proportional to $\partial_v \ell_v$ vanishes (since here $k>1$) and one obtains the relation 
\be  \label{relationg}
  g'(v) J_0(v) - J_1(v)  = 0 \;.
\ee 
On the other hand, taking the derivative of the first equation in \eqref{system} w.r.t. $v$ gives
\be  \label{chi_sec}
- N^{-\frac{(n+1)k+n}{k+n}}  \partial^2_v \tilde \chi(v,N) =  
g'(v) J_1(v) - J_2(v)   =  
 \frac{J_1(v)^2}{J_0(v)}  - J_2(v) \;,
\ee 
where we have used \eqref{relationg} to eliminate $g'(v)$. 
Taking $v=0$ one finds the expression for the variance
\be
{\rm Var} \left(\tilde {\cal L}_N\right) = N^{\frac{k+n-n k}{k+n}}  \left( 
    J_2(0) - \frac{J_1(0)^2}{J_0(0)} \right) \;.  
\ee 
Using that upon rescaling one has 
\be 
J_p(0) = 2 \frac{a_k}{k} \left(\frac{\ell_0}{2}\right)^{ n(\frac{1}{k}-1) + 1} u_n^{\frac{1}{k}-1} I_p \quad , \quad 
I_p = \int_0^1 dz f\left(\frac{\ell_0}{2} z\right)^p (1- z^n)^{\frac{1}{k}-1}
\ee 
One recovers the formula \eqref{var_sr} for the variance given in the text. 
\\

{\bf Higher order cumulants}. It possible to compute the higher cumulants by taking successive derivatives of 
Eq. \eqref{chi_sec} w.r.t. $v$. This procedure for a general $f(y)$ becomes very tedious for higher cumulants,
and he we illustrate it only for the case of the third cumulant. The main difficulty is to eliminate $\ell_v$
and its derivatives iteratively. One can obtain $\partial_v \ell_v$ from \eqref{relationg} $g'(v) = J_1(v)/J_0(v)$ 
and the definition of $g(v)$ in \eqref{system}. This leads to
\be \label{partialell}
\partial_v \ell_v 
= \frac{2 (J_1(v)/J_0(v) - f( \frac{\ell_v}{2} )) }{ n u_n (\frac{\ell_v}{2})^{n-1} + v f'(\ell_v/2)} \;.
\ee 
To take derivatives of Eq. \eqref{chi_sec} w.r.t. $v$ we also need to compute the derivatives of $J_p(v)$.
To do this, we first need to rescale the integrals appearing in $J_p(v)$ in order to avoid evaluating
the integrand for $y=\ell_v/2$ (where it diverges). Hence one uses the rescaled form (obtained by performing the change of variable $y = \ell_v z/2$)
\be  \label{scaled_Jp}
J_p(v) =  \frac{a_k}{k} \ell_v \int_{0}^{1} dz \left[f\left(z \frac{\ell_v}{2}\right)\right]^p \bigg[ u_n \left(\frac{\ell_v}{2}\right)^n (1-z^n) + 
v \left(f\left(\frac{\ell_v}{2}\right) - f\left(z \frac{\ell_v}{2}\right)\right) 
\bigg]^{\frac{1}{k} - 1 } \;,
\ee  
where we used that $f(y)$ is even in $y$. Using this expression we can now take derivatives of Eq. \eqref{chi_sec} w.r.t. $v$
and use iteratively \eqref{partialell} to eliminate the derivatives of $\ell_v$. This leads to the result for the 
third cumulant given in \eqref{third_cumul_sr}. 

\subsubsection{Power law linear statistics $f(y)=f_m |y|^m $}

We now specialize to power law linear statistics of the form
\be 
f(y) = f_m |y|^m \quad , \quad {\rm with} \quad m \geq 1 \;.
\ee 
In that case one can use rescaling to simplify the equations. 
By performing the change of variable $y= \frac{\ell_v}{2} z$ in the second equation in \eqref{system} 
one obtains 
\be 
 \ell_v^{- (1 + \frac{n}{k})} = {a_k}  \left( \frac{u_n}{2^n}\right)^{1/k} g(\hat v) \quad , \quad 
 g(\hat v) := \int_0^1 dz \left[ 1 - z^n + \hat v \left(1 - z^m \right) \right]^{1/k} \quad, \quad \hat v = v \frac{f_m}{u_n} \left(\frac{\ell_v}{2}\right)^{m-n} \;. \label{scaled_g} 
\ee 
Upon the same change of variable the first equation in \eqref{system} gives
\bea  
&& - N^{-\frac{(n+1)k+n}{k+n}} \partial_v \tilde \chi(v,N) = \ell_v^{1 + \frac{n}{k}+ m} 
\frac{f_m}{2^m}  {a_k}  \left( \frac{u_n}{2^n}\right)^{1/k} 
h(\hat v) \quad, \quad h(\hat v) = \int_0^1 dz z^m \left[ 1 - z^n + \hat v \left(1 - z^m \right) \right]^{1/k}  \label{scaled_h}
\eea
We can now eliminate $\ell_v$ between \eqref{scaled_g} and \eqref{scaled_h} leading to a parametric representation of 
$\partial_v \tilde \chi(v,N)$ as a function of $v$, via the variable $\hat v$
\bea
&& v = \frac{u_n}{f_m} 2^{m-n} \left( \frac{u_n}{2^n}\right)^{\frac{m-n}{k+n}} \hat v \, \left[a_k g(\hat v)\right]^{\frac{k(m-n)}{k+n}} \\
&& - \frac{1}{N^{\frac{(n+1)k+n}{k+n}}} \partial_v \tilde \chi(v,N) = \frac{f_m}{2^m} a_k^{- \frac{m k}{k+n} }
\left( \frac{u_n}{2^n} \right)^{- \frac{m}{k+n} }
\left[g(\hat v)\right]^{- 1 - \frac{m k}{k+n} } h(\hat v)\label{new} 
\eea  
One now trades the derivatives w.r.t. $v$ with those w.r.t. $\hat v$ using
\be  \label{der} 
 \partial_v = \frac{f_m}{u_n} 2^{n-m} \left( \frac{u_n}{2^n}\right)^{\frac{n-m}{k+n}}  
 \bigg[  (a_k g(\hat v))^{\frac{k(m-n)}{k +n}}  (1 + \frac{k(m-n)}{k+n} \hat v \frac{g'(\hat v)}{g(\hat  v)})  \bigg]^{-1} \,  \partial_{\hat v} 
\ee 
The cumulants are thus obtained from the relation [see Eq. (\ref{sumsum})]
\be 
\langle \tilde{\cal L}_N^q \rangle_c = N^{- \frac{q n k}{k+n} } (-1)^q \partial_v^q \tilde \chi(v,N) \big \vert_{v=0} \;.
\ee 
Using \eqref{new} and \eqref{der} one then obtains the formula given in \eqref{formulacum} above. 

To evaluate successively the cumulants for $q \geq 2$ from formula  \eqref{formulacum} we see that we need 
the derivatives of the functions $g(\hat v)$ and $h(\hat v)$ at $\hat v=0$. Although there does not appear
to be useful closed forms for the integrals which define these functions in Eqs. \eqref{scaled_g} and \eqref{scaled_h},
they can be expanded in powers of $\hat v$ and the coefficients of this expansion can be computed explicitly
as follows
\bea 
\int_0^1 dz \left[(1-z^n) + \hat v (1-z^m)\right]^{1/k} = \sum_{j \geq 0} { 1/k \choose j} \hat v^j 
\int_0^1 dz
(1-z^n)^{\frac{1}{k}-j} (1-z^m)^j \;,
\eea 
where the generic term reads 
\bea 
\int_0^1 dz
(1-z^n)^{a-j} (1-z^m)^j  = \sum_{r=0}^j { j \choose r} (-1)^r \int_0^1 dz
(1-z^n)^{\frac{1}{k}-j} z^{m r} = \sum_{r=0}^j { j \choose r} (-1)^r 
\frac{\Gamma(1+\frac{1}{k}-j) \Gamma(\frac{1+m r}{n}) }{n \Gamma(1+ \frac{1}{k} - j + \frac{1+m r}{n} ) } \;.
\eea 
This leads to the equation \eqref{gseries} above. Similarly the series for the function $h(\hat v)$
is obtained from 
\bea 
\int_0^1 dz z^m \left[(1-z^n) + \hat v (1-z^m)\right]^{1/k} = \sum_{j \geq 0} { 1/k \choose j} \hat v^j 
\int_0^1 dz z^m
(1-z^n)^{\frac{1}{k}-j} (1-z^m)^j 
\eea 
where the generic term reads 
\bea 
\int_0^1 dz z^m
(1-z^n)^{\frac{1}{k}-j} (1-z^m)^j  &=& \sum_{r=0}^j { j \choose r} (-1)^r \int_0^1 dy
(1-z^n)^{\frac{1}{k}-j} z^{m (r+1)} \\
&=& \sum_{r=0}^j { j \choose r} (-1)^r 
\frac{\Gamma(1+\frac{1}{k}-j) \Gamma(\frac{1+m (r+1)}{n}) }{n \Gamma(1+ \frac{1}{k} - j + \frac{1+m (r+1)}{n} ) } \nonumber
\eea 
This leads to the equation \eqref{hseries} above. We recall that in all the above formula we have $k>1$
since here we are studying the short range Riesz gas.




Let us display here the first three cumulants for general $n,m > 1$. For convenience let us define
\be 
\langle \tilde{\cal L}_N^q \rangle_c \approx N^{1 - \frac{(q-1) n k}{k+n} } (-1)^{q+1} A_q W_{n,m,q} 
\ee 
where $A_q$ is the amplitude given in \eqref{def_Aq}. Then one finds for the first moment
\bea 
 W_{n,m,1}  = \frac{ 
{ \frac{1}{n}+\frac{1}{k} \choose \frac{1}{n} }^{\frac{k m}{k+n}+1}  }
{ (m+1) { \frac{1}{k}+\frac{m+1}{n} \choose \frac{1}{k} } 
} 
\eea 
For the second cumulant we find
\bea W_{n,m,2}  = 
{ \frac{1}{n}+\frac{1}{k} \choose \frac{1}{n} }^{\frac
   {k (2 m-n)}{k+n}} 
\frac{\Gamma \left(\frac{1}{n}+1+\frac{1}{k}\right)
   \left(\frac{\Gamma
   \left(\frac{1}{n}+\frac{1}{k}\right) \Gamma
   \left(\frac{m+1}{n}\right)^2}{\Gamma
   \left(\frac{m+1}{n}+\frac{1}{k}\right)^2}-\frac{\Gamma \left(\frac{1}{n}\right) \Gamma \left(\frac{2
   m+1}{n}\right)}{\Gamma \left(\frac{2
   m+1}{n}+\frac{1}{k}\right)}\right)}{\Gamma
   \left(\frac{1}{n}\right)^2} \;.
\eea  
For the third cumulant we find
\bea 
&&W_{n,m,3}  
=  { \frac{1}{n}+\frac{1}{k} \choose \frac{1}{n} }^{\frac{3 k m-2 k n+k+n}{k+n}} \frac{\Gamma \left(1+\frac{1}{k}\right) }{n^2}  
\\
&& \times 
 \left(\frac{(k
   (3 m-2 n+2)+2 n) \Gamma
   \left(\frac{1}{n}+\frac{1}{k}\right)^2 \Gamma
   \left(\frac{m+1}{n}\right)^3}{k \Gamma
   \left(\frac{1}{n}\right)^2 \Gamma
   \left(\frac{m+1}{n}+\frac{1}{k}\right)^3}-\frac{3 n
   \Gamma \left(\frac{1}{n}+\frac{1}{k}\right) \Gamma
   \left(\frac{2 m+1}{n}\right) \Gamma
   \left(\frac{m+1}{n}\right)}{\Gamma
   \left(\frac{1}{n}\right) \Gamma
   \left(\frac{m+1}{n}+\frac{1}{k}\right) \Gamma
   \left(\frac{2 m+1}{n}+\frac{1}{k}-1\right)}+\frac{n
   \Gamma \left(\frac{3 m+1}{n}\right)}{\Gamma
   \left(\frac{3
   m+1}{n}+\frac{1}{k}-1\right)}\right) \;. \nonumber 
\eea 
The explicit evaluation of these first three cumulants is given in Table \ref{table_CM} for the Caloger-Moser model (i.e. $k=2$ and $U(y) = u_2 y^2$) for different values of $m=1,2, \cdots, 5$. 

\subsubsection{Counting statistics}\label{subsubsec:counting}

Let us present two equivalent methods to obtain the cumulants of the number of
particles ${\cal N}_L$ in the interval $y \in [-L/2,L/2]$. 

{\bf First method}. One starts from the parametric system \eqref{system} and insert $f(y)=\theta(\frac{L}{2}-y)\theta(y+\frac{L}{2})$.
One then rescales $y = \frac{\ell_v}{2} z$ and obtains the equivalent system
\bea \label{systemnew1} 
 1 &=& 2 a_k u_n^{\frac{1}{k}} \left(\frac{\ell_v}{2}\right)^{1 + \frac{n}{k} } \left(  f_{\frac{1}{k}}\left( \frac{L}{\ell_v} , \hat v  \right)  
+ g_{\frac{1}{k}}\left(\frac{L}{\ell_v} \right)  \right) \quad , \quad \hat v = \frac{v}{u_n} \left(\frac{2}{\ell_v}\right)^n\\
 - \beta N^{-\frac{(n+1)k+n}{k+n}}  \partial_v \tilde \chi(v,N) &=& 2 a_k u_n^{\frac{1}{k}} \left(\frac{\ell_v}{2}\right)^{1 + \frac{n}{k} } 
 f_{\frac{1}{k}}\left( \frac{L}{\ell_v} , \hat v \right) = 1 - 2 a_k  u_n^{\frac{1}{k}} \left(\frac{\ell_v}{2}\right)^{1 + \frac{n}{k} } g_{\frac{1}{k}}(\frac{L}{\ell_v}) \label{systemnew2} 
\eea 
where $\tilde g_b(r)$ was defined in \eqref{def_gtilde} and we further introduced the functions
\bea 
&& g_{\frac{1}{k}}(r) = \int_r^1 dz (1- z^n)^{1/k} = \tilde g_{n,\frac{1}{k}}(1) - \tilde g_{n,\frac{1}{k}}(r) 
\\
&& f_{\frac{1}{k}}(r,\hat v) = \int_0^r dz (1- z^n- \hat v)^{1/k} = (1-\hat v)^{\frac{1}{n} + \frac{1}{k} } \tilde g_{\frac{1}{k}} \left( \frac{r}{(1-\hat v)^{1/n} } \right) \;,
\eea  
Note that the first derivative of $\tilde g_{\frac{1}{k}}(r)$ is simple and the derivative of $f_{\frac{1}{k}}(r,\hat v)$
w.r.t $\hat v$ can be written as
\be 
 \partial_{\hat v} f_{b}(r,\hat v) = - b \int_0^r dz (1- z^n- \hat v)^{b-1}  = - b f_{b-1}(r,\hat v) 
\quad , \quad f_{b}(r,0) = \tilde g_{b}(r)  \;,
\ee 
which are all well defined integrals (since $\hat v$ is vanishingly small and $r<1$).
Finally, setting $v=0$ in Eq. \eqref{systemnew1} gives the relation
which determines $\ell_0$ as
\be 
1 = 2 a_k u_n^{\frac{1}{k}} \left(\frac{\ell_0}{2}\right)^{1 + \frac{n}{k} } \tilde g_{\frac{1}{k}}(1) 
\ee 
which is equivalent to \eqref{ell0} given above. 

Since $\tilde g_b(r)$ is given explicitly in terms of an hypergeometric function, see Eq. \eqref{def_gtilde}, 
the functions $f_{b}(r,\hat v)$ and $g_b(r)$ are all explicit. The calculation proceeds as follows. 
One first obtains from Eq.~\eqref{systemnew1}, using Mathematica, 
an expansion of $\ell_v$ to any given order in powers of $v$ around $\ell_0$. 
Inserting this expansion in (\ref{systemnew2}) gives the expansion of
the generating function $\partial_v \chi(v,N)$ in powers of $v$, leading
to the cumulants. Note that this method is very efficient and fast if one specifies some values for $k$ and $n$, e.g.
for the CM model $k=n=2$. 
\\

{\bf Second method}. In that case we start instead from the formula \eqref{chi_sec} for the second derivative of the generating function.
The integrals $J_p(v)$ which appear in this formula are defined in \eqref{def_Jp}. In the present case
where $f(y)=\theta(\frac{L}{2}-y)\theta(y+\frac{L}{2})$ they read
\bea 
&& J_1(v)=J_2(v)= J(v) :=  \int_{0}^{L/2} dy   \, [\left(\frac{\ell_v}{2}\right)^n - u_n y^n - v  ]^{\frac{1}{k}-1 } \;, \\
&& J_0(v) = J(v) + I(v) \quad , \quad  I(v) = \int_{L/2}^{\ell_v/2} dy   \, [\left(\frac{\ell_v}{2}\right)^n - u_n y^n   ]^{\frac{1}{k}-1 } \;.
\eea 
Performing a rescaling $y = \frac{\ell_v}{2} z$ we can rewrite \eqref{chi_sec} as
\be  \label{chi_sec2}  
 \beta N^{-\frac{(n+1)k+n}{k+n}}  \partial^2_v \tilde \chi(v,N) = \frac{2 a_k}{k} \frac{ I(v) J(v) }{I(v) + J(v) } 
= \frac{2 a_k}{k} u_n^{\frac{1}{k}-1} \left(\frac{\ell_v}{2}\right)^{1 + n(\frac{1}{k}-1) }  
\frac{ g_{\frac{1}{k}-1}(\frac{L}{\ell_v}) f_{\frac{1}{k}-1}(\frac{L}{\ell_v},\hat v) }
{ g_{\frac{1}{k}-1}(\frac{L}{\ell_v}) + f_{\frac{1}{k}-1}(\frac{L}{\ell_v},\hat v) } 
\quad , \quad \hat v = \frac{v}{u_n} \left(\frac{2}{\ell_v}\right)^n \;.
\ee 
To take derivatives of this formula w.r.t. $v$ we use Eq. \eqref{partialell} which gives $\frac{d \ell_v}{dv}$. In terms of rescaled integrals
it reads
\be \label{dlvdv}
 \frac{d \ell_v}{dv} =  \frac{2}{n} \frac{J(v)  }{ J_0(v)  (\frac{\ell_v}{2})^{n-1} } 
= \frac{2}{n u_n (\frac{\ell_v}{2})^{n-1} } \frac{  f_{\frac{1}{k}-1}(\frac{L}{\ell_v},\hat v) }
{ g_{\frac{1}{k}-1}(\frac{L}{\ell_v}) + f_{\frac{1}{k}-1}(\frac{L}{\ell_v},\hat v) }  \;.
\ee  
We can now obtain the cumulants, which, as we recall, are given by the derivatives
\be  \label{cumgen}
\langle {\cal N}_L^q \rangle_c = (-1)^q \beta^{1-q} N^{-\frac{q n k}{k+n}}\partial^q_v \tilde \chi(v,N)|_{v=0} 
\ee 
The variance is readily obtained by setting $v=0$ in \eqref{chi_sec2} and leads to
the formula \eqref{var_FCS_res} given above. 
To obtain the higher cumulants we use the chain rule
\be 
\partial_v = \frac{1}{u_n} [ (\frac{2}{\ell_v})^n + v  \frac{d}{dv} (\frac{2}{\ell_v})^n  ]
\partial_{\hat v} + \frac{d \ell_v}{dv} \partial_{\ell_v} 
\ee 
where $\frac{d \ell_v}{dv}$ is given in \eqref{dlvdv}. In the simplest case of the third cumulant
we can evaluate this operator at $v=0$ and it simplifies into
\be 
\partial_v|_{v=0} = \frac{1}{u_n}  (\frac{2}{\ell_0})^n \left( \partial_{\hat v}|_{\hat v=0} + \frac{\ell_0}{n} \frac{  \tilde g_{\frac{1}{k}-1}(\frac{L}{\ell_0}) }
{ \tilde g_{\frac{1}{k}-1}(1) }   \partial_{\ell_v}|_{\ell_v=\ell_0} 
\right) 
\ee 
Using this formula to take the derivative w.r.t. $v$ of \eqref{chi_sec2}, together with \eqref{cumgen} 
we obtain the result \eqref{third_cumul_FCS} displayed above for the third cumulant. We have checked that the first and second method presented here give the same results.
The calculations beyond the third cumulant become more involved, and are detailed
in the Appendix \ref{App:FCS}.

\section{Conclusion} \label{sec:conclusion}

In this paper we have studied the one-dimensional Riesz gas at equilibrium in an even external potential which
leads to an average density which is symmetric around the origin and supported on a single interval. 
We have focused on monomial external potentials of the form $|x|^n$. 
We have considered both the
case of long range (LR) interactions, i.e. $\sim 1/|x|^k$ with $-2<k<1$, and the
case of short range (SR) interactions corresponding to $k>1$. The physics is rather
different in these two cases. Indeed, in the first case ($-2<k<1$) we have scaled the external potential 
so that the support remains of order unity in $x$. In the second case ($k>1$) one needs
to scale the spatial coordinate with the number of particles $N$ as $x = N^{k/(k+n)} y$ so that the support of
the equilibrium density remains of order unity in the $y$ coordinate. 

Our main interest has been on the
linear statistics, i.e., the fluctuations of the observable ${\cal L}_N= \sum_{i=1}^N f(x_i)$ in the large $N$ limit,
where $f(x)$ is a smooth even function. In the LR case we have obtained explicit formulae for the cumulants of ${\cal L}_N$ of arbitrary order $q$
when $f(x)$ is a monomial of even {integer} power. These cumulants are found to scale as $N^{2-q}$. 
We have compared our result for the second cumulant to a recent work \cite{Beenakker_riesz} which addresses the case $-1<k<1$
for a larger class of functions $f(x)$, and found agreement with our formulae.
Since the SR case behaves differently, we have considered rescaled
linear statistics of the form $\tilde {\cal L}_N= \sum_{i=1}^N f(y_i)$ in the large $N$ limit.
We have obtained explicit formulae for the cumulants of $\tilde {\cal L}_N$ in 
two cases: (i) for $f(y)$ an arbitrary smooth even function, we have
obtained the first three cumulants. Our method allows in
principle to obtain higher ones, but the calculation becomes very tedious;  
(ii) for $f(y) \sim |y|^m$ with $m>1$ we have obtained a general iterative formula for arbitrary
cumulants in an external potential of the form $|y|^n$. In that case, the cumulant of order $q$ is found to scale as $N^{1- \frac{(q-1) n k}{k+n}}$. 
As an application of the later case we have specialized our formulae to the 
Calogero-Moser model ($k=n=2$) and displayed explicit expressions for the first three cumulants 
for integer $m$ up to $m=6$ (see Table I). Finally, again for the general SR Riesz gas, we have computed the cumulants
of the number of particles ${\cal N}_L$ in the interval $[-L/2,L/2]$. We have obtained explicit
formula for the cumulants of order $q=2,3,4$ (extending a result for $q=2$ in \cite{Riesz_FCS}).
These cumulants are singular when $L$ approaches the edge of the support $\ell_0$. One thus expects
that a separate edge regime occurs when $L \approx \ell_0$, with a crossover from the
bulk regime to this edge regime. Although this edge regime remains to be explored, we
have obtained here the scaling forms of (i) the cumulant generating function and (ii) the corresponding large deviation
rate function associated to the PDF of ${\cal N}_L$ upon approaching $\ell_0$ from the bulk.

Our results in the LR case present an interesting generalization of previously known formulae for (i) the log-gas $k \to 0$
and (ii) the Coulomb gas $k=-1$. In the first case (i) a general formula was available for the second
cumulant for an arbitrary smooth function $f(x)$. Interestingly it depends only on the support
of the density, and not on the precise form of the potential, and encompasses several examples
of interest in random matrix theory. As we have shown, our formula for the second cumulant for general $k$ and 
for even monomial potentials reproduces these
known results in the limit $k \to 0$. In addition, we have also obtained explicit formula for the higher cumulants of
arbitrary order which, to our knowledge, were not known, even for the log-gas. For the 
Coulomb gas (ii) we have shown that all cumulants obtained in previous works are correctly reproduced
in the limit $k \to -1$. Remarkably, in these works, it was found 
that all cumulants of order $3$ and higher are determined only by the edge of the support \cite{UsCoulomb}. 
We find here that this feature does not hold for general $k$ and is thus a peculiar feature of the
Coulomb interaction. 


From our results for the cumulant generating function of the linear statistics, we have also obtained, via a Legendre transform, some information of the large deviation form of the probability distribution of the linear statistics, not too far in the tails. Further in the tails 
we have observed that there is a transition which is similar to the "evaporation transition" found recently in the
case of the log-gas \cite{Valov} which we recover here in the limit $k \to 0$. 

We end up by mentioning some open 
problems left for future works. It would be interesting
to further study the large deviation form of the distribution of the linear statistics of the LR Riesz gas.
Indeed, in addition to the above mentioned "evaporation transition"
other transitions are likely to occur, as found in the case
of the log-gas \cite{Valov}, where the support of the optimal density splits
into several disjoint intervals. Extension of the present results to higher dimensions
is also a great challenge, although recently it was achieved in the case of the Coulomb interaction \cite{UsCoulomb}.
Finally, there is currently some interest about the dynamics of Riesz gas
\cite{Mallick_riesz,SerfatyDynamics1,SerfatyDynamics2,flack2023out,PLDRankDiff}, but large deviations for linear and counting statistics
remain to be studied in this context.

  \section*{Acknowledgments}

{We thank Satya N. Majumdar for very useful and stimulating discussions, especially in the earlier stages of this work,
and for sharing a calculation which is summarized in Appendix \ref{app:cm}.} We acknowledge support from 
ANR grant ANR-23-CE30-0020-01 EDIPS. We thank LPTMS for hospitality.
PLD also thanks KITP for hospitality, 
supported by NSF Grants No. NSF PHY-1748958 and PHY-2309135. 
    

\bigskip

\begin{appendix}

\section{Cumulants of the linear statistics in the limit of the log-gas $k=0$} \label{app_log_gas}
\label{app:loggas} 

\subsection{Second cumulant}

The case of the log-gas is obtained in the limit $k \to 0$, with $J k = 1$ fixed. 
We can compare our results with the predictions of Ref. \cite{Vivo2014} where the variance of
${\cal L}_N= \sum_i f(x_i)$ is computed for various random matrix ensembles. The result
can be expressed only as a function of the support of the equilibrium density
(without specifying the details of the confining potential). Here we choose
a quadratic potential $V(x)=x^2/2$ (which corresponds to the Gaussian ensemble of random matrix theory) in which case one finds that 
the size of the support is $\ell_0=2 \sqrt{2}$.
The predictions from Ref. \cite{Vivo2014} are then 
\bea  \label{vivo} 
&& l_0^{-2m}\beta {\rm Var} \left(\sum_i \frac{x_i^m}{m} \right) = \frac{1}{\pi^2 } \int_0^{+\infty} dk k \tanh(\pi k) |\tilde f(k)|^2 \\
&& \tilde f(k) = \frac{1}{m} \int_{-\infty}^{+\infty} e^{i k x} \left[T(e^{x})\right]^m dx \quad, \quad T(z) = \frac{z\,\lambda_- + \lambda_+}{z+1} \;,
\eea
where the equilibrium measure has a support on $[\lambda_-, \lambda_+]$. Here we have rescaled
by $\ell_0$ hence we must use $\lambda_\pm = \pm 1/2$, in which case
$\tilde f(k) = \frac{(-1)^m}{2^m m} \int_{-\infty}^{+\infty} e^{i k x} (\tanh(x/2))^m$. 
On the other hand, our prediction for the variance \eqref{var1} reads, with $p=k+1=1$ in the present case,
 \be \label{var1app} 
 l_0^{-2m}\beta {\rm Var} \left(\sum_i \frac{x_i^m}{m} \right) =  c_{m,m} \;,
\ee
where $J k$ must be set to unity. We have checked that for $m=2$ both Eqs. \eqref{vivo} and \eqref{var1app} agree with the value $1/(256)$ and
for $m=4$ both formulae give $9/131072$. Remarkably, we also find that the continuation of our formula
for $c_{m,m}$ to $m$ odd integer also agree with the prediction \eqref{vivo}. However, as noted in the
text, this property holds only for $p=k+1=1$.
Note that there are some other works which have derived an equivalent for the formula \eqref{vivo}, see
e.g. Refs. \cite{Lambert}. Also we have checked that formula (3.6) of \cite{Beenakker_riesz} 
agrees for $p=q=m$ for integer values $m=1,2,3,4$ with the prediction from \eqref{vivo}. 

\subsection{Higher cumulants by the rescaling method for $m=n$}

We can check the log-gas case $k=0$, $p=1$, for any $n=m$ using rescaling method. 
Starting from
\be  \label{ZNU}
Z_{N,U} = \int_{\mathbb{R}^N} dx_1 \dots dx_N  \, e^{- \beta ( N \sum_i U(x_i) - \frac{1}{2}  \sum_{i \neq j} \log |x_i-x_j| ) } 
\ee 
Let us consider the case where $U(x)=u_n |x|^n$ and $f(x)=u_n |x|^n$. Using 
that $\chi(s,N)= \log Z_{N,U+ f s/\beta}  - \log Z_{N,U}$ and rescaling space 
as $x \to x/(1+s/\beta)^{1/n}$ one finds the following result, which is exact for any $N$
\be 
- \frac{1}{N^2} \chi(s,N) = \frac{\beta}{2 n} \left(1 + \frac{1}{N}\right) \log(1+s/\beta) \;.
\ee 
This gives the exact expression of the cumulants
\be \label{exactcum} 
\langle {\cal L}_N^q \rangle_c =  \beta^{1-q} \frac{1}{N^{q-2}} \frac{(q-1)!}{2 n}  (1 + \frac{1}{N})  \;.
\ee 
We can compare with our prediction from Eqs. \eqref{higherc}, \eqref{condnorm} 
and \eqref{c22nm}. For $p=1$ and $m=n$ we find that $a(\lambda) = \lambda^{1+n}/n$,
and $c_{nn}/c_n^2=1/(2 n)$, which leads to
\be 
 \langle {\cal L}_N^q \rangle_c 
 = \beta^{1-q} \frac{1}{N^{q-2}} \frac{1}{2 n} \left[ \left(  \frac{\lambda^{1+n}}{n}  \frac{d}{d\lambda} \right)^{q-2} \lambda^{2 n} \right]_{\lambda=1}
 = \frac{(q-1)!}{2 n} \frac{1}{N^{q-2}} 
\ee 
and thus agrees with the result \eqref{exactcum} from the rescaling method to leading order at large $N$.

\section{Comparison with the covariance calculations of Ref. \cite{Beenakker_riesz}} 
\label{app:Beenakker} 

In Ref. \cite{Beenakker_riesz}, C. W. J. Beenakker computes the covariance of some observables for the Riesz gas
for $-1<k<1$, using a slightly different method. The model is the same, as defined by our equation \eqref{model}.
The dictionary between the notations in his
paper and in the present work are as follows
%
%

\begin{eqnarray} \label{corres}
\begin{array}{|c||c|c|c|c|c|c|c|c|}
\hline
{\rm Ref.}~[88] & s & s_+ = {(s+1)}/{2} & s_-={(s-1)}/{2} & q & p & q=p & L & X \\
\hline
{\rm Here} & k=p-1 & p/2 & p/2-1 & m_1 & m_2 & m & \ell & x \\
\hline 
\end{array}
\end{eqnarray}
The observables are also slightly different. In \cite{Beenakker_riesz} the external potential (that we call here $U(x)$) is not specified,
since it is shown there 
that the covariances do not depend on it. 
They only depend on the size
of the support of the equilibrium density $[a,b]$, which is chosen to be $X \in [0,L]$ there
and $x \in [-\ell_0/2,\ell_0/2]$ here, where for clarity we use the letter $X$ to denote
the variable $x$ in \cite{Beenakker_riesz}.
In that work a general formula, Eq. (3.2) there, for the covariance of $\sum_i f(X_i)$ 
and $\sum_i g(X_i)$ is obtained, but it is not explicit. In the case 
$f(X)=X^{m_1}$ and $g(X)=X^{m_2}$ explicit formulae are given, in Eq. (3.3) there. 
In the present work, we give a formula for the variance of the symmetric centered
monomials $f(x)=g(x)=u_m x^m$. It is possible to compare the results using the relation
\be \label{relationxX}
x= \ell_0 \left(\frac{X}{L}-1/2\right) \;,
\ee 
as we discuss below. In the Appendix of \cite{Beenakker_riesz}, the following
integral equation for $S(x)$ on the interval $[a,b]$ is considered
\be \label{integral_eq_app}
J |k| \int_{a}^{b} dx' \frac{{\rm sgn}(x'-x)}{|x-x'|^{k+1}} S(x') = g'(x) \;.
\ee 
Note that this equation is identical to our Eq. \eqref{integral_eq} for the equilibrium density, setting $g(x)=U(x)$.
It is recalled that for $-1<k<1$ it has two special solutions \cite{Sonin,stanley_riesz}
\bea 
&& S_-^{g}(z)  = - A (z-a)^{p/2-1} \frac{d}{dz} \int_z^b dt (t-a)^{1-p} (t-z)^{p/2} 
 \frac{d}{dt} \int_a^t dz' (z'-a)^{p/2} (t-z')^{p/2-1} g'(z') \label{Sm}\\
&& S_+^{g}(z)  = A (z-a)^{p/2} \frac{d}{dz} \int_z^1 dt (t-a)^{1-p} (t-z)^{p/2-1} 
 \frac{d}{dt} \int_a^t dz' (z'-a)^{p/2-1} (t-z')^{p/2} g'(z') \;, \label{Sp}
\eea 
where $A = - \frac{1}{J |k|} \frac{2 \sin(\pi p/2)}{\pi p B(p/2,p/2)}$ 
is denoted $C_1$ in \cite{Beenakker_riesz}. For the same range $-1<k<1$ there exists a non trivial
solution to the homogeneous equation (i.e. for $g'(x)=0$ in Eq. (\ref{integral_eq_app})) 
which reads
\be \label{S0}
S_0(z) = ((z-a) (b-z))^{p/2-1} \;,
\ee 
which is indeed normalizable for $p>0$. The relation between the functions $S_-^g$, $S_+^g$ and $S_0$, in the case of a polynomial function $g(x)$, is discussed in Ref. \cite{Beenakker_stack}.

Let us recall that in the present paper our method to compute 
the cumulants consists in: (i) specifying a potential $U(x)$, (ii) shifting
this potential $U(x) \to U(x) + s f(x)$ (iii) determining the optimal
density associated to this shifted potential, using Eq. \eqref{integral_eq_app} 
with $g(x) \to U(x) + s f(x)$, and also determining its support $[-\ell_s/2,\ell_s/2]$
using the normalization condition. In the latter stage we use the solution $S_-^{U + s f}(z)$ given in (\ref{Sm}) with
the substitution $g \to U + s f$.
The method of Ref. \cite{Beenakker_riesz} (which only aims to compute the second moments)
is quite different since, there, the support $[0,L]$ is fixed (and the potential
is not specified). In that case however the solution is constructed from
a linear combination of $S_+^{g}$ and $S_0$. We have checked that the two 
methods give identical results at the level of the second moments (whenever
it was possible to compare the results). This seems a priori quite non
trivial. A similar property was used in the Coulomb gas case $k=0$ \cite{Vivo2014}. 

Let us now compare the results of Ref. \cite{Beenakker_riesz} with our formula for the
variance \eqref{var1}. Using the formula (3.3) in that paper for the covariance 
of $X_{m_1}=\sum_i X_i^{m_1}$ and $X_{m_1}=\sum_i X_i^{m_2}$ taking into
account \eqref{relationxX} 
we see that the following identity should hold for positive integer $m$
\bea \label{shift}
&&\ell_0^{-2 m}  \, \beta {\rm Var}\left( \sum_i \frac{x_i^m }{m}\right) =  \beta \,  {\rm Var}\left( \sum_i \frac{\left(\frac{X_i}{L} - \frac{1}{2}\right)^m }{m}\right) \nonumber \\
&& =\frac{\beta}{m^2}\sum_{k_1=0,k_2=0}^m {m \choose k_1} {m \choose k_2} \left(- \frac{1}{2}\right)^{2 m-k_1-k_2} \frac{\langle X_{k_1} X_{k_2} \rangle}{L^{k_1+k_2}} =    c_{m,m} \ell_0^{p-1} 
\frac{m}{m+p-1}  = \eqref{var1} \;.
\eea
We have checked this non trivial identity for $m=2,4,6$ and
$p \in ]-1,2[$. Remarkably, we found that this formula also holds for $p<0$, if one uses the natural
analytical continuations to $p<0$ of the covariance formula given in Eq. (3.5) of Ref. \cite{Beenakker_riesz},
while the method used there does not allow to reach that regime.


\section{Explicit calculation of the coefficient $c_m$ for even integer $m$}
\label{app:cm} 

In this Appendix, we consider the following Sonin integral, which coincides with one in the first line of \eqref{definitioncnm} for even integer $m$ 
\begin{equation}
c_m = A\, \int_0^1 dz z^{p/2-1} \frac{d}{dz} \int_z^1 dt t^{1-p} (t-z)^{p/2}
 \frac{d}{dt} \int_0^t dz' (z')^{p/2} (t-z')^{p/2-1}  (z'-1/2)^{m-1}  \, ,
\label{cm_def}
\end{equation}
where $A$ is given in \eqref{def_A}.
We show below that $c_m$ can be exactly computed and is given by the
formula
\begin{equation}
c_m= (-1)^{m-1}\, \frac{2^{-m}}{J \,|k|\, \pi}\,  \sin\left(\frac{\pi p}{2}\right)\,  B\left(\frac{p}{2},
\frac{p}{2}\right)\, 
{}_2F_1\left[1-m, 1+\frac{p}{2}, 1+p, 2\right]\, .
\label{cm_exact.1}
\end{equation}
For $m$ even integer one can check (using \cite{wolfram}) that it agrees with the formula given in the text in \eqref{formulacm}. 

The proof of the result in Eq. (\ref{cm_exact.1}) proceeds via the following steps.
For convenience, in this Appendix we denote 
\begin{equation}
n'=m-1\,.
\label{n_def}
\end{equation}
\vskip 0.2cm

\noindent{\bf {Step 1:}} \hskip 0.2cm Let us first define the rightmost integral that appears in Eq. (\ref{cm_def}) 
\begin{equation}
I_{n'}(t)=
\int_0^t dz' (z')^{p/2} (t-z')^{p/2-1}  (z'-1/2)^{n'}\, ,
\label{In_def}
\end{equation}
where we keep the $p$-dependence in $I_{n'}(t)$ implicit for ease of notations.
Rescaling $z'=y\, t$ in Eq. (\ref{In_def}) 
and 
expanding $(yt-1/2)^{n'}$ in a binomial series we get, performing the integral over $y$ term by term,
\begin{equation}
I_{n'}(t)= \sum_{k=0}^{n'} {n' \choose k}\, (-1/2)^{n' -k}\, t^{p+k}\, B\left(\frac{p}{2}+k+1,\frac{p}{2}\right)\, .
\label{Int.2}
\end{equation}
Finally, taking derivative with respect to $t$ gives
\begin{equation}
\frac{dI_{n'}(t)}{dt}= \sum_{k=0}^{n'} {n' \choose k}\, (-1/2)^{n'-k}\, (p+k)\, t^{p+k-1}\, B\left(\frac{p}{2}+k+1,\frac{p}{2}\right)\, .
\label{der_Int.1}
\end{equation}

\vskip 0.3cm

\noindent{\bf {Step 2:}} \hskip 0.2cm
We now substitute Eq. (\ref{der_Int.1}) in the expression for $c_m$ in Eq. (\ref{cm_def}) and
consider the integral over $t$, namely we define
\begin{equation}
J_{n'}(z)= \int_z^1 dt\, t^{1-p}\, (t-z)^{p/2}\, \frac{dI_{n'}(t)}{dt}= \sum_{k=0}^{n'} {n' \choose k}\,
(-1/2)^{n'-k}\, (p+k)\,  B\left(\frac{p}{2}+k+1,\frac{p}{2}\right)\, \int_z^1 dt\, t^k\, (t-z)^{p/2}\, .
\label{Jn_def}
\end{equation}
Note that we need the derivative $dJ_{n'}(z)/dz$ in Eq. (\ref{cm_def}). 
If we take a derivative with respect to $z$ directly in Eq. (\ref{Jn_def}), the term coming
from integrating over $z$ in the lower limit of the integral will diverge if $p<0$. Hence
it is more convenient to first do 
the integral over $t$ in (\ref{Jn_def}) once by parts and then take the derivative with respect to $z$.
Integrating once by parts we have
\begin{eqnarray}
\label{Jn.1}
\int_z^1 dt\, t^k\, (t-z)^{p/2} &= & \, t^k\, \frac{(t-z)^{p/2+1}}{(p/2+1)} \Big|_{t=z}^{t=1} - 
\frac{k}{(p/2+1)}\, \int_z^1 dt\,
t^{k-1}\, (t-z)^{p/2+1} \nonumber \\
&= & \frac{1}{(p/2+1)}\, \left[(1-z)^{p/2+1} - k \, \int_z^1 dt\,
t^{k-1}\, (t-z)^{p/2+1} \right] \, .
\end{eqnarray}
Note that there is no contribution from the lower limit $t=z$, since it vanishes for $p>-1$. Now, taking
a derivative with respect to $z$ gives
\begin{equation}
\frac{dJ_{n'}(z)}{dz}= \sum_{k=0}^{n'} {n' \choose k}\,
(-1/2)^{n'-k}\, (p+k)\, B\left(\frac{p}{2}+k+1,\frac{p}{2}\right)\, \left[- (1-z)^{p/2} + k\,  
\int_z^1 dt\, t^{k-1}\, (t-z)^{p/2} \right]\, .
\label{der_Jn.1}
\end{equation}

\vskip 0.3cm

\noindent{\bf {Step 3:}} \hskip 0.2cm
We next substitute $dJ_{n'}(z)/dz$ from (\ref{der_Jn.1}) into Eq. (\ref{cm_def}) to write
\begin{eqnarray}
\label{cm.2}
c_m &= & A\, \int_0^1 dz\, z^{p/2-1}\, \frac{dJ_{n'}(z)}{dz}   \\
&= & A \sum_{k=0}^{n'} {n'\choose k}\,
(-1/2)^{n'-k}\, (p+k)\, B\left(\frac{p}{2}+k+1,\frac{p}{2}\right)\, \left[ - B\left(\frac{p}{2}+1,\frac{p}{2}\right) 
+k\, \int_0^1 dz\, z^{p/2-1}\, \int_z^1 dt\, 
t^{k-1}\, (t-z)^{p/2}\right]\, . \nonumber
\end{eqnarray}
Now to perform the double integral in the last term over the region $t\in [z,1]$ and $z\in [0,1]$,
it is convenient to instead consider the same region but with $z\in [0,t]$ and $t\in [0,1]$. This gives
\begin{eqnarray}
\label{double_int.1}
\int_0^1 dz\, z^{p/2-1}\, \int_z^1 dt\,
t^{k-1}\, (t-z)^{p/2}&= & \int_0^1 dt\, t^{k-1}\, \int_0^t dz \, z^{p/2-1}\, (t-z)^{p/2}\, \nonumber \\
&=& \frac{1}{p+k}\, B\left(\frac{p}{2}+1, \frac{p}{2}\right)\, .
\end{eqnarray}
Note that in arriving to the second line from the first, we made the rescaling $z= t\, u$ which makes
the integral over $u$ straightforward.

\vskip 0.3cm

\noindent{\bf {Step 4:}} \hskip 0.2cm
Finally, substituting the result from Eq. (\ref{double_int.1}) into the expression for $c_m$
in the second line of (\ref{cm.2}) and simplifying, we get a compact expression
\begin{equation}
c_m=- A\,p\,  B\left(\frac{p}{2}+1,\frac{p}{2}\right)\,  
 \sum_{k=0}^{n'} {n'\choose k}\,
(-1/2)^{n'-k}\, B\left(\frac{p}{2}+k+1,\frac{p}{2} \right) \, .
\label{cm.3}
\end{equation}
One can evaluate this sum over $k$ explicitly in terms of an hypergeometric function, namely,
\begin{equation}
\sum_{k=0}^{n'} {n'\choose k}\,
(-1/2)^{n'-k}\, B\left(\frac{p}{2}+k+1,\frac{p}{2} \right)
= (-1)^{n'} \sqrt{\pi}\, 2^{-n'-p}\, \frac{\Gamma\left(\frac{p}{2}\right)}{\Gamma\left(\frac{1+p}{2}\right)}\,
{}_2F_1\left[-n', 1+\frac{p}{2}, 1+p, 2\right]\, .
\label{iden.1}
\end{equation}
We substitute this result in (\ref{cm.3}), use the expression for $A$ in (\ref{def_A}) and simplify to get
\begin{equation}
c_m= (-1)^{n'}\, \frac{1}{J\, |k|\, \pi}\, \sin\left(\pi\, \frac{p}{2}\right)\, \sqrt{\pi}\, 2^{-n'-p}\,
\frac{\Gamma\left(\frac{p}{2}\right)}{\Gamma\left(\frac{1+p}{2}\right)}\, 
{}_2F_1\left[-n', 1+\frac{p}{2}, 1+p, 2\right]\, .
\label{cm.4}
\end{equation}
This formula can be further simplified by using the doubling formula of Gamma function, namely
\begin{equation}
\Gamma(p)= \frac{1}{\sqrt{\pi}}\, 2^{p-1}\, \Gamma\left(\frac{p}{2}\right)\, \Gamma\left(\frac{1+p}{2}\right)\,.
\label{doubling_g.1}
\end{equation}
Using this identity in (\ref{cm.4}) and $n'=m-1$,  we obtain precisely the expression given in \eqref{cm_exact.1}.



\section{Double sum formula for the coefficients $c_{n,m}$ for $n,m$ even integer}
\label{app:cnm} 

One can use the derivation in the previous Appendix to obtain a double sum representation for
$c_{n,m}$ for $n,m$ even integer. From the definition \eqref{definitioncnm} we see that
we can setting $m \to n$ in the previous Appendix, and 
insert a factor $\frac{1}{m} (z-\frac{1}{2})^m$ in the Step 4 of the previous calculation. 
Using the binomial formula $(z-\frac{1}{2})^m = \sum_{\ell=0}^m (-1/2)^{m-\ell} z^\ell$
we see that for each value of $\ell$ we can simply insert $z^\ell$ in \eqref{cm.2}.
Performing the same steps then leads to 
\begin{equation}
c_{n,m}=- \frac{A}{m} \,  \sum_{k=0}^{n-1} {n-1 \choose k}\,
(-1/2)^{n-1-k}\, (p+k) B\left(\frac{p}{2}+k+1,\frac{p}{2} \right) 
\sum_{\ell=0}^m  {m \choose \ell}\, (-1/2)^{m-\ell}\
\frac{p+\ell}{k+p+\ell}  B\left(\frac{p}{2}+1,\frac{p}{2} + \ell \right) 
\label{cnm.1}
\end{equation}
where $A = - \frac{1}{J |k|} \frac{2 \sin(\pi p/2)}{\pi p B(p/2,p/2)}$.
For $n,m$ even integers we have checked that this formula agrees with the formula given in the text 
\eqref{c22nm}. 

\section{Simplest short range case: $f(y)=|y|$ for the Calogero-Moser model ($k=2$ and $n=2$)}

One can test our results for the cumulants for the short range model on the simplest case which corresponds to
the Calogero-Moser model, i.e. $n=2$ with $u(y) = y^2 $, $m=1$, with $f(y)=|y|$ and $k=2$. 
%
In that case the system \eqref{system} becomes 
\bea \label{system2} 
&& - N^{-2}  \partial_v \tilde \chi(v,N) =  
2 a_2 \int_{0}^{\ell_v/2} dy y \, \bigg[ g(v) - \left( 
y^2 + v y \right ) 
\bigg]^{\frac{1}{2}} \\
&& 1 = 
2 a_2  \int_{0}^{\ell_v/2} dy \bigg[ g(v)  - \left( 
y^2 + v y \right ) 
\bigg]^{\frac{1}{2}} \quad , \quad g(v) =  \left(\frac{\ell_v}{2}\right)^2 + v \frac{\ell_v}{2} \;,
\eea 
where $[a_2=3 J \zeta(2)]^{-1/2}$. In this case, these integrals can be explicitly evaluated. 
The normalisation condition (second line in \eqref{system2}) reads
\be \label{firstline} 
1= \frac{1}{4} a_2 \left(2 \left(\ell_v+v\right){}^2 \sin
   ^{-1}\left(\frac{\sqrt{\frac{\ell_v}{\ell_v+v}}}{\sqrt{2}}\right)-v \sqrt{\ell_v
   \left(\ell_v+2 v\right)}\right) 
\ee 
from which we can extract the series expansion of $\ell_v$
\bea \label{ellseries}
&& \ell_v = \ell_0 + \sum_{n \geq 1} b_n a_2^{\frac{n-1}{2}} v^n \quad , \quad \ell_0^2 = \frac{8}{\pi a_2}  \\
&& b_1 =  \frac{2}{\pi }-1 \quad , \quad b_2= \frac{1}{\sqrt{2}
   \pi ^{3/2}} \quad , \quad b_3 =  -\frac{1}{24} \quad , \quad b_4 = \frac{\pi ^2-3}{24 \sqrt{2} \;.
   \pi ^{5/2}} \label{direct_n2}
\eea 
The integral in the first line of \eqref{system2} can be computed leading to
\bea  
- N^{-2}  \partial_v \tilde \chi(v,N) &=&
\frac{1}{24} a_2 \left(\sqrt{\ell_v \left(\ell_v+2 v\right)} \left(4 v \ell_v+2
   \ell_v^2+3 v^2\right)-6 v \left(\ell_v+v\right){}^2 \sin^{-1}\left(\frac{\sqrt{\frac{\ell_v}{\ell_v+v}}}{\sqrt{2}}\right)\right) \\
   &=& \frac{1}{12} a_2 \ell_v \left(\ell_v+2 v\right) \sqrt{\ell_v \left(\ell_v+2
   v\right)}-\frac{v}{2}
\eea  
where the second line is obtained using \eqref{firstline}. Using the above expansion \eqref{ellseries} one finds
\bea &&  \label{seriesapp} - N^{-2}  \partial_v \tilde \chi(v,N) \\
&& 
= \frac{4 \sqrt{2}}{3 \pi ^{3/2} \sqrt{a_2}}+\left(\frac{4}{\pi
   ^2}-\frac{1}{2}\right) v-\frac{\left(\left(\pi ^2-12\right)
   \sqrt{a_2}\right) v^2}{2 \left(\sqrt{2} \pi
   ^{5/2}\right)}-\frac{\left(\left(\pi ^2-8\right) a_2\right) v^3}{3 \pi
   ^3}+\frac{\left(240-40 \pi ^2+3 \pi ^4\right) a_2^{3/2} v^4}{192
   \sqrt{2} \pi ^{7/2}}+O\left(v^5\right) \nonumber \;.
\eea  
Using that $\langle \tilde {\cal L}_N^q \rangle_c = (-1)^q N^{-q} \partial_v^q \chi(v,N)|_{v=0}$
one can check that this series reproduces the first three cumulants, as given in the Table \ref{table_CM}
(see caption there using $u_2=f_2=1$ and $A_q=(2 a_2)^{\frac{q-2}{2}}$.
\\

\noindent{\bf Remark}. Note that here we did a direct calculation by determining the modified support size $\ell_v$. 
Alternatively one can obtain the series \eqref{seriesapp} using, as in the main text, the formula given in (\ref{chi_sec})
in terms of the integrals $J_p(v)$ defined in the text in Eq. (\ref{scaled_Jp}).
These integrals can be explicitly computed as
\bea
&& J_0(v) = 2 a_2 \sin
   ^{-1}\left(\frac{\sqrt{\frac{\ell_v}{\ell_v+v}}}{\sqrt{2}}  
   \right) \simeq a_2 \frac{\pi}{2} - a_2 v/\ell_0 \;, \\
&& J_1(v) = \frac{1}{2} a_2 \left(\sqrt{\ell_v \left(\ell_v+2 v\right)}-2
   v \sin
   ^{-1}\left(\frac{\sqrt{\frac{\ell_v}{\ell_v+v}}}{\sqrt{2}}
   \right)\right) \;,
\\
&& 
J_2(v) = a_2 \left(\frac{1}{4} \left(2 v \ell_v+\ell_v^2+3 v^2\right)
   \sin
   ^{-1}\left(\frac{\sqrt{\frac{\ell_v}{\ell_v+v}}}{\sqrt{2}}
   \right)-\frac{3}{8} v \sqrt{\ell_v \left(\ell_v+2
   v\right)}\right) \;.
\eea 
Using $ - N^{-2}  \partial^2_v \tilde \chi(v,N) = 
\frac{J_1(v)^2}{J_0(v)} - J_2(v)$ one recovers correctly the derivative 
of the series in \eqref{seriesapp} above.



\section{Calculation of the higher cumulants for the counting statistics}\label{App:FCS}

To go beyond the third cumulant of ${\cal N}_L$ we need to make the method more systematic. We define the 
variable $s=L/\ell_v$, and set $b=\frac{1}{k}-1$. Then it is easy to see that the
general formula for the cumulants $q \geq 3$ can be written as
\bea 
&& \langle {\cal N}_L^q \rangle_c \simeq N^{1-(q-1)\frac{nk}{k+n}} \beta^{1-q}\,(-1)^q \frac{2 a_k}{k} 
(\frac{\ell_0}{2} )^{1 + n (\frac{1}{k}+1-q) } u_n^{1 + \frac{1}{k} - q} \\
&& \times r^{1 + n (b+3-q)} 
 \left( \partial_{\hat v} - \frac{\tilde g_{b}(r) }
{ \tilde g_{b}(1)} \frac{r}{n} \partial_{s}   \right) 
\bigg[ s^n 
 \left( \partial_{\hat v} + \frac{f_{b}(s,\hat v) }
{ g_{b}(s) + f_{b}(s,\hat v) } ( - \frac{s}{n} \partial_{s} -  \hat v \partial_{\hat v} ) \right) \bigg]^{q-3} 
\left( s^{- ( 1 + n b)  }  
\frac{ g_{b}(s) f_{b}(s,\hat v) }
{ g_{b}(s) + f_{b}(s,\hat v) } \right) \bigg|_{s=r,\hat v=0} \;,\nn 
\eea  
with $r= L/\ell_0$. One can perform iteratively the derivatives using mathematica and applying the following rules
\bea  
&&  \partial_{\hat v} f_{b}(a,\hat v)  = - b f_{b-1}(a,\hat v) \quad , \quad \partial_{a} f_{b}(a,\hat v) = (1-a^n - \hat v)^b 
\quad , \quad f_{b}(a,0) = \tilde g_{b}(a) \\
&& g_b'(a)=- (1-a^n)^b \quad , \quad \tilde g_b'(a)= (1-a^n)^b \quad , \quad g_b(a)= \tilde g_b(1)- \tilde g_b(a) 
\eea 
This leads to formula for the higher cumulants. For the fourth cumulant we obtain for generic $n,k>1$
\bea  
&& \langle {\cal N}_L^4 \rangle_c \simeq   \frac{2 a_k}{k} 
(\frac{\ell_0}{2} )^{1 + n (\frac{1}{k}-3) }  u_n^{\frac{1}{k} - 3} \bigg[ 
-
\frac{r^2 h_1(r) \left(4 h_1(r)-1\right)
   \left(1-r^n\right)^{\frac{2}{k}-2}}{n^2 \tilde g_{\frac{1}{k}-1}(1) } \\
   && - \frac{\tilde g_{\frac{1}{k}-1}(1)}{k^2 n^2} 
   (h_1(r)-1) \bigg(2 (k-1)^2 n^2
   (h_1(r)-1) h_2(r){}^2-(k-1) (2 k-1)
   n^2 (h_1(r)-1) h_3(r)  \\
   && +h_1(r){}^3 (-2 k
   n+k+n) (k (-n)+k+n)+(k-1) n \left(h_1(r)-1\right)
   h_2(r) h_1(r) (k (5 n-3)-3 n)\bigg) \\
&& + \frac{r \left(1-r^n\right)^{\frac{1}{k}-2} }{k n^2}
   \bigg(h_1(r){}^3 \left( (k (5 n-3)-3 n) r^n-9 k
   n+3 k+7 n\right)+h_1(r){}^2 \big(6 (k-1) n
   h_2(r) \left(r^n-1\right)\\
   && +(-3 k n+2 k+2 n) r^n+8
   k n-2 k-7 n \big)-(k-1) n h_1(r) \left(7 h_2(r)
   \left(r^n-1\right)+2\right)+(k-1) n h_2(r)
   \left(r^n-1\right)\bigg) \bigg] \nn 
\eea 
where the functions $h_j(r)$ have been defined in \eqref{def_gtilde}. For $k=n=2$ this formula reduces 
to the result \eqref{CMCS4} for the fourth cumulant of ${\cal N}_L$ for the CM model. 
\\

{\bf Conjecture for the behavior of the cumulants near $r=1$}. To obtain the leading behavior for $r \to 1$ it turns out that 
one only needs to compute, for $q \geq 2$
\bea 
&& \langle {\cal L}_N^q \rangle_c \simeq (-1)^q \frac{2 a_k}{k} 
(\frac{\ell_0}{2} )^{1 + n (\frac{1}{k}+1-q) }  u_n^{1 + \frac{1}{k} - q} 
 \left( -  \frac{1}{n} \partial_{s}   \right)^{q-2} g_{b=\frac{1}{k}-1}(s)  \bigg|_{s=r} \\
 && = (-1)^q \frac{2 a_k}{k} 
(\frac{\ell_0}{2} )^{1 + n (\frac{1}{k}+1-q) } u_n^{1 + \frac{1}{k} - q} 
k n^{\frac{1}{k}+1-q} \frac{ \Gamma(1+ \frac{1}{k}) }{\Gamma(3-q+ \frac{1}{k})} (1-r)^{\frac{1}{k}+2 - q} \;.
\eea  
Indeed, one can check that for $s$ close to $1$ and $\hat v < 1-s$ also small one has
\bea  
&& g_b(s) = \tilde g_b(1)- \tilde g_b(s) \simeq \frac{n^b}{1+b} (1-s)^{1+b} \;, \\
&& f_b(s,\hat v) = \tilde g_b(1) - \frac{n^b}{1+b} (1-s- \frac{1}{n} \hat v)^{1+b} \;,
\eea  
which leads to great simplifications. In addition one can check that
the operator $\partial_{\hat v} - \frac{1}{n} \partial_s $ does not act on
$f_b$ to leading order, and that the term $\hat v \partial_{\hat v}$
is subdominant.

\section{Distribution of ${\cal N}_L$ for $k>1$ and the limit $r = \frac{L}{\ell_0} \to 1$}\label{app:prob}

\subsection{Scaling form of the PDF of ${\cal N}_L$ in the limit $r \to 1$ using the results of Ref. \cite{Riesz_FCS}}

The PDF of ${\cal N}_L$ for the short range Riesz gas in a quadratic well ($n=2$) was studied
in \cite{Riesz_FCS}. It was shown that at large $N$ it takes the large deviation form (in the notations of the present paper)
\be  \label{LDF_FCS}
{\cal P}({\cal N}_L=c N ) \sim \exp\left(  - \beta N^{1 + \frac{n k}{k+n}} \Phi(c, r) \right)
\ee 
where $r=L/\ell_0$ and the function $\Phi(c,r)$ was obtained in \cite{Riesz_FCS} for $n=2$ and $u_n=1/2$. 
However the form above (\ref{LDF_FCS}) is supposed to hold for any $n>1$ with an $n$-dependent rate functions. 
In that paper the
function is called $\Phi(c,w)$ where 
$w=r \frac{\ell_0}{2}$. We will also use the notations $\ell= \lambda \frac{\ell_0}{2}$
and $\bar \ell= \bar \lambda \frac{\ell_0}{2}$ as compared to that paper. With these notations 
\be \label{Phidef} 
\Phi(c, r)= \frac{\ell_0^2}{8} \frac{k+2}{3 k + 2} \left( (1-c)^{\frac{3 k+2}{k+2}} {\cal H}(\frac{r}{\lambda}) 
+ c^{\frac{3 k+2}{k+2}} {\cal J}(\frac{r}{\bar \lambda})  - 1 \right) 
\ee 
in the regime $c > \bar c(r)= r^{1 + \frac{2}{k}}$ relevant here. Note that in that regime
$\lambda,\bar \lambda > r$. The functions which appear in \eqref{Phidef} are defined as
\bea  
&& {\cal H}(h) = \left(1- I(h^2, \frac{1}{2}, 1 + \frac{1}{k})\right)^{- \frac{2 k}{k+2} } 
+ \frac{h (1-h^2)^{1 + \frac{1}{k}} (2 k^2)}{ (k+1) (k+2) B( \frac{1}{2} , 1 + \frac{1}{k} ) }
\left(1- I(h^2, \frac{1}{2}, 1 + \frac{1}{k})\right)^{- \frac{3 k+2}{k+2} } \label{def_H} \\
&& {\cal J}(h) = I(h^2, \frac{1}{2}, 1 + \frac{1}{k})^{- \frac{2 k}{k+2} } 
- \frac{h (1-h^2)^{1 + \frac{1}{k}} (2 k^2)}{ (k+1) (k+2) B( \frac{1}{2} , 1 + \frac{1}{k} ) }
I(h^2, \frac{1}{2}, 1 + \frac{1}{k})^{- \frac{3 k+2}{k+2} }  \;, \label{def_J}
\eea  
where 
\be 
I(r^2 , \frac{1}{2} , \frac{1}{k} +1 ) = \frac{ \tilde g_{\frac{1}{k}}(r)}{\tilde g_{\frac{1}{k}}(1)} \;, \label{def_I}
\ee 
in terms of the function $\tilde g_b(r)$ defined in \eqref{def_gtilde}. Here 
$\lambda$ and $\bar \lambda$ depend on $c$ and $r$ and are
determined as solutions of 
\bea  
&& c = \bar \lambda^{ \frac{k+2}{k} }  \, I( (\frac{r}{\bar \lambda})^2 , \frac{1}{2} , \frac{1}{k} +1) \;,\label{eq_lambda} \\
&& 1-c = \lambda^{ \frac{k+2}{k} } \, ( 1- I( (\frac{r}{ \lambda})^2 , \frac{1}{2} , \frac{1}{k} +1)) ) \;. \label{eq_lambda_bar}
\eea  
The typical value (i.e., the average value) of ${\cal N}_L$ corresponds to $c=c^*=c^*(r)$ such that $\lambda=\bar \lambda=1$ leading to
\be 
\frac{\langle {\cal N}_L \rangle}{N} \simeq  c^* =  \frac{\tilde g_{\frac{1}{k}}(r)}{\tilde g_{\frac{1}{k}}(1)} \;.
\ee 

The function $\Phi(c,r)$ is analytic around $c^*$ since it is analytic for $c>\bar c(r)$ with $c^*>\bar c(r)$, see Fig. 2
in \cite{Riesz_FCS}. As shown there, there is a phase transition across the line $c=\bar c(r)$, which
is not of interest here.
By expanding $\Phi(c,r)$ around $c=c^*$ one can in principle obtain the cumulants of ${\cal N}_L$.
The calculation is however extremely tedious and has been done only for the second cumulant in \cite{Riesz_FCS}. 
The method we use here in the main text to compute the cumulants is different, but the two methods should
agree as we have checked for the second cumulant. 
\\

{\bf Regime $r \to 1$}. In that regime we will check that the following ansatz given in the main text [see Eq. (\ref{def_PHI_txt})] holds
\be  \label{leading_PHI}
1-c = \gamma (1- r)^{\frac{1}{k}+1} \quad , \quad 
\Phi(c,r) \simeq (1-r)^{\frac{1}{k}+2} \phi(\gamma) \quad , \quad 
\ee 
Here we will obtain the function $\phi(\gamma)$ by two different methods (i) directly from the
above expression for $\Phi(c,r)$ given above for $n=2$, specializing the calculation
to $k=2$ for simplicity (ii) upon Legendre transform by a summation of the cumulants 
obtained in the present paper. We find agreement between the two methods.

Let us start with (i). To obtain $\phi(\gamma)$ we use the expansion of the function $I$ in Eq. (\ref{def_I}), which
to lowest order reads 
\be 
\frac{ \tilde g_{\frac{1}{k}}(r)}{\tilde g_{\frac{1}{k}}(1)}  = I(r^2 , \frac{1}{2} , \frac{1}{k} +1 )  \simeq 1 - \gamma^* (1-r)^{\frac{1}{k}+1} \quad , \quad 
\gamma^* 
= \frac{n^{\frac{1}{k}} \Gamma
   \left(\frac{1}{n}+1+\frac{1}{k}\right)}{\Gamma
   \left(2+\frac{1}{k}\right) \Gamma
   \left(1+\frac{1}{n}\right)}
\ee 
from which one obtain the expansions of the functions ${\cal H}$ and ${\cal J}$.a
From Eq. \eqref{eq_lambda} we can then extract the value of $\lambda$ and we find, to leading order 
\be 
1-\lambda = (1-r) (1- (\frac{\gamma}{\gamma^*})^{\frac{k}{k+1}} ) + O((1-r)^2) \;. \label{exp_lambda}
\ee 
Similarly from \eqref{eq_lambda_bar} one can extract the leading order of the expansion for $\bar \lambda$ which reads
\be 
1- \bar \lambda \simeq \frac{k}{k+2} (\gamma - \gamma^*) (1-r)^{1 + \frac{1}{k}}
\ee 
and one can check that $\tilde \lambda, \lambda>r$. Substituting the expansions of $\lambda, \bar \lambda$ in Eqs. \eqref{def_H}
and \eqref{def_J} and specializing to $k=2$ for simplicity we obtain to the desired order in the expansion in $1-r$
\bea 
\gamma^{\frac{3 k+2}{k+2}}
(1- r)^{\frac{(k+1)(3 k+2)}{k (k+2)}}
{\cal H}(\frac{r}{\lambda})  = 2 \gamma (1-r)^{3/2} + \frac{\gamma}{10} ( - 40 + 3 (6 \pi)^{2/3} \gamma^{2/3} ) (1-r)^{5/2} + \dots 
\eea 
\bea 
(1- \gamma (1-r)^{\frac{1}{k}+1})^{\frac{3 k+2}{k+2}} {\cal J}(\frac{r}{\bar \lambda})  - 1 = 
- 2 \gamma (1-r)^{3/2} + \frac{64 \sqrt{2}}{15 \pi} (1-r)^{5/2} + O((1-r)^3) \dots  
\eea 
To obtain these expressions we have expanded $\lambda$ in \eqref{exp_lambda} to one more order, while 
it turns out that we could simply set $\bar \lambda = 1$. Putting together
we see that the leading term cancel and that one finally obtains \eqref{leading_PHI} with 
\be 
\phi(\gamma) = \frac{\ell_0^2}{16}  \gamma^* \left(  \frac{8}{5} - 4 \frac{\gamma}{\gamma^*} + \frac{12}{5} (\frac{\gamma}{\gamma^*})^{5/3} \right) \;.
\ee 
This expression agrees with the general one given in the main text \eqref{phig}, setting $k=n=2$, $u_n=1/2$ and using $\ell_0^2 = 8/(\pi a_2 u_n^{1/2})$ with $a_k$ given in Eq. (\ref{ak_SR}).

\subsection{Scaling form of the PDF of ${\cal N}_L$ in the limit $r \to 1$ by resumming the cumulants}

We first recall the relation between the PDF and the generating function $\tilde \chi(v,N)$ computed in this paper.
From its definition in \eqref{def_chi_SR}, and using the large deviation form for ${\cal P}({\cal N}_L)$ at large $N$ given 
in \eqref{LDF_FCS}, we can write 
\be 
e^{ \tilde \chi(v,N)  } = \sum_{{\cal N}_L=0}^\infty  {\cal P}({\cal N}_L) \exp( - v N^{\frac{n k}{k+n}} {\cal N}_L )
\propto \int dc \exp( - v N^{1+\frac{n k}{k+n}} c  - \beta N^{1+\frac{n k}{k+n}} \Phi(c,r) ) \;.
\ee 
A saddle point argument then leads to the relation
\be \label{Leg} 
\tilde \chi(v,N) = - N^{1+\frac{n k}{k+n}} \min_{c \in [0,1]} ( \beta\, \Phi(c,r) + v \,c ) \;.
\ee 
Inserting the scaling form \eqref{leading_PHI} we find that near $r=1$ this relation becomes (setting $c= 1- \gamma (1-r)^{1/k+1}$)
\be \label{Leg2} 
\tilde \chi(v,N) = - N^{1+\frac{n k}{k+n}} v - N^{1+\frac{n k}{k+n}} (1-r)^{\frac{1}{k}+2} \min_{\gamma \in \mathbb{R}} ( \beta\, \phi(\gamma) - \gamma \frac{v}{1-r} ) 
\ee  
which predicts that as $r \to 1$, the generating function $\tilde \chi(v,N)$ takes a scaling form as a function of the parameter $v/(1-r)$
with a scaling function related to $\phi(\gamma)$ by a Legendre transformation.

To determine $\phi(\gamma)$ we
now explicitly compute $\tilde \chi(v,N)$ for $r \to 1^-$ by summing the cumulants of ${\cal N}_L$.
Using the expression given in \eqref{cumul_r1} we see that it indeed takes a scaling form
as a function of $v/(1-r)$ which reads
\bea 
&& \tilde \chi(v,N) = \sum_q \frac{(-1)^q}{q!} N^{\frac{n k q}{n+k}} v^q \langle {\cal N}_L^q \rangle_c 
\simeq - N^{1 + \frac{n k}{k+n} }v + N^{1 + \frac{n k}{k+n} } (1- r)^{\frac{1}{k} + 2} B \chi(\hat u) 
\\
&& 
B = \beta 2 a_k (\frac{\ell_0}{2} )^{1 + n (\frac{1}{k}+1) } \frac{k^2}{(1+k)(1+2 k)} (n u_n)^{1 + \frac{1}{k}} 
\quad , \quad \hat u= \frac{1}{\beta n u_n} \left(\frac2{\ell_0}\right)^{n} \frac{v}{1-r}    \\
&&  \chi(\hat u) = \sum_{q \geq 1} \frac{1}{q!} 
 \frac{ \Gamma(3+ \frac{1}{k}) }{\Gamma(3-q+ \frac{1}{k})}\,\hat u^q =  (1+\hat u)^{2 + \frac{1}{k}} - 1 \label{chihatu} \;.
\eea 

We now need to perform the Legendre inversion of Eq. \eqref{Leg2} to obtain $\phi(\gamma)$ from $ \chi(\hat u)$ near $r=1$. One 
obtains
\be 
\beta \phi(\gamma) = B \max_{\hat u} (\frac{\gamma}{\gamma^*} (2 + \frac{1}{k})  \hat u -  \tilde \chi(\hat u) ) 
\ee 
with
\be 
\gamma^* = (2 + \frac{1}{k}) \frac{B}{ \beta (n u_n) (\frac{\ell_0}{2})^n} 
= n^{1/k} \frac{\Gamma(1 + \frac{1}{k} + \frac{1}{n})}{\Gamma(2 + \frac{1}{k}) \Gamma(1 + \frac{1}{n})}
\ee 
where in the last identity we used \eqref{ell0}. The Legendre inversion using the expression for $\chi(\hat u)$ in \eqref{chihatu} then leads to
our final result for the function $\phi(\gamma)$
\bea 
\phi(\gamma)= \beta^{-1} B \left( 1 - (2 + \frac{1}{k}) \frac{\gamma}{\gamma^*} + (1 + \frac{1}{k})  (\frac{\gamma}{\gamma^*})^{\frac{1+2 k}{1+k} } \right) 
\eea 
which coincides with the result given in the main text in \eqref{phig}.


\section{Comparison of the PDF rate function $\Psi(\Lambda)$ with Ref. \cite{Valov} in the case of the log-gas} \label{app:psilog}

In the special case of the log-gas, $k \to 0$ with $J |k|=1$, i.e. $p=1$,
our results can be compared to those of a recent work \cite{Valov} on the Gaussian random matrix ensemble.
That case corresponds to setting $p=1$ and $n=2$ in our formula, and considering the potential $U(x)=x^2/2$,
which corresponds to $u_2=1/2$. From the normalization condition \eqref{condnorm} one then has $n u_n c_n \ell_0^2 = 1$ and $c_2=1/8$
and one finds $\ell_0=2 \sqrt{2}$. 
We can compare with the case of
no spectral gap (single support of the optimal density), i.e. with 
their formulae (43)-(44) upon the 
identification (with $f_m=1$) 
\be 
\lambda = \frac{\ell_s}{\ell_0} \to \sqrt{B/2} \quad , \quad \Lambda \to A \quad , \quad s  \to \Lambda \quad , \quad f_m \to 1 \;.
\ee 
Beware that the same symbol $\Lambda$ denotes different quantities in the different papers.
Consider our formulae \eqref{psi_main} and \eqref{ltilde} for $p=1$ and $n=2$ 
\be \label{La}
  \Lambda = m  \frac{c_{m,m}}{c_m} \ell_0^m ( \lambda^m - (1- \frac{c_m c_{2 m}}{c_2 c_{m,m}} ) \, \lambda^{2+m} ) 
  = \frac{c_{m,m}}{c_m} \ell_0^m \lambda^m 
 (1- \frac{m-2}{m+2}  \, \lambda^{2} ) 
\ee  
Setting  $p=1$ in \eqref{formulacm} and \eqref{c22nm} one obtains 
\be 
\frac{c_{m,m}}{c_m} = \frac{2^{-m} \Gamma \left(\frac{m+1}{2}\right)}{\sqrt{\pi }
   m^2 \Gamma \left(\frac{m}{2}\right)}  \quad , \quad c_m = \frac{2^{-m}}{\sqrt{\pi}} \frac{ \Gamma(\frac{1+m}{2})}{\Gamma(1 + \frac{m}{2}) } \;.
\ee 
If one uses these values, Eq. \eqref{La} becomes, in the notations of \cite{Valov}
\be 
\Lambda \to  A=  m B^{m/2} (1 - \frac{m-2}{m+2} \frac{B}{2}) \frac{\Gamma(\frac{m+1}{2}) }{m^2 \sqrt{\pi} \Gamma(\frac{m}{2}) }
\ee 
where the r.h.s. coincides with (43) in \cite{Valov} setting $m f_m=1$.
Using the normalization condition one finds 
\bea  
&& \Psi'(\Lambda) = \frac{1}{m f_m c_m \ell_0^m} \tilde \Psi'(\tilde \Lambda) 
= \frac{1}{m  c_m \ell_0^m} \lambda^{-m} (\lambda^{2} - 1)  \to \Lambda = \frac{\sqrt{\pi} \Gamma(\frac{m}{2}) }{4  \Gamma( \frac{1+m}{2} ) } B^{-m/2} (B-2) 
\eea  
which the formula (44) in \cite{Valov}. Hence the rate function coincides in the two papers. 
Note that our derivation holds for $m$ even integer while the derivation in \cite{Valov} is for any real $m$
(within the region of a single support). 

Our prediction for the transition also coincides with the discussion around Eq. (48) 
in \cite{Valov} since we find from \eqref{lambdac} (setting $p=1$)
\be 
\lambda_c^{2} = \frac{m}{m-2}  = \frac{B^*}{2} \;,
\ee 
where $B^*$ denotes the location of the evaporation transition in \cite{Valov}. 
\\

{\bf Coulomb gas}. As discussed in the main text, in the case of the Coulomb gas $k=-1$, $p=0$, $J=1$, the PDF large deviation rate function $\Psi(\Lambda)$ 
was computed in \cite{UsCoulomb} for any space dimension $d$ (see Appendix B) and given in a parametric form in Eq. (B5) there. We use these
formula for polynomial $U(x)$ and $f(x)$ with arbitrary $m,n>1$. 
One can check that (up to a factor $\beta$) our parametric system 
\eqref{systemPsi} and \eqref{psi_main}, 
reproduces correctly
Eq. (B5) in \cite{UsCoulomb} for $p=0$ and $d=1$, using the coefficients given in 
\eqref{cnCG} as well as the normalisation condition \eqref{condnorm}
with the identification $R_s=\lambda \ell_0/2$. Note that
we have used \eqref{systemPsi} with the first line replaced by \eqref{ltilde}, 
which thus appears to be valid for any $n,m$ (at least in the case $p=0$).
Note that \eqref{cnCG} leads to $2^n 
\frac{c_{m,m} c_n}{c_m^2} = 2^m \frac{c_{m,m}}{c_m}  = \frac{m-1}{m (2 m-1) }$.
Hence the above evaporation transition extends to the Coulomb gas, and occurs at the
critical value $\lambda_c = \frac{m-1}{m-n}$, leading to a critical
value for $R_s$, 
$R_{s_c}=\lambda_c \ell_0/2$ and for $s$, $s_c= - \Psi'(\Lambda_{\max})$ with
\bea \label{lambda_max}
\Lambda_{\max} = f_m 2^{-m} \frac{m-1}{2m-1} \ell_0^m \frac{n-1}{m+n-1} \left( \frac{m-1}{m-n}\right)^{\frac{m}{n-1}} \;.
\eea
One can check that this corresponds to
the point $s_c$ is such that $dR_s/ds|_{s=s_c}=+\infty$
since the numerator in the function $A(R_s)$ defined in (B1) of in \cite{UsCoulomb} vanishes
for $R_s=R_{s_c}$.





\bigskip

\newpage

\end{appendix}


\begin{thebibliography}{100}
\bibitem{footnote1}
The rule to restore the $u_n$ factors is to do $a_k \to a_k/u_n^{1/n}$ and $\ell_0 \to u_n^{1/n} \ell_0$.
\end{thebibliography}
\end{document}